\UseRawInputEncoding
\documentclass[journal,twocolumn]{IEEEtran}
\usepackage{stfloats}
\usepackage{setspace}
\usepackage{algorithm}
\usepackage{algorithmic}
\usepackage{amsfonts}
\usepackage{dsfont}
\usepackage{enumerate}
\usepackage{cite,array}
\usepackage{graphicx}
\usepackage{amssymb}
\usepackage{lipsum}
\usepackage[cmex10]{amsmath}
\usepackage{cite}
\usepackage{cases}
\usepackage{mathrsfs}
\usepackage{color, soul, framed}
\usepackage[colorlinks,linkcolor=blue,citecolor=blue,urlcolor=blue]{hyperref}
\usepackage{graphicx}
\usepackage{float}
\usepackage{bm}
\usepackage{arydshln}
\usepackage{graphicx}
\usepackage{epstopdf}
\newtheorem{theorem}{Theorem}
\newtheorem{corollary}{Corollary}
\newtheorem{lemma}{Lemma}
\newtheorem{proposition}{Proposition}
\newtheorem{definition}{Definition}
\newtheorem{remark}{Remark}

\newtheorem{assumption}{Assumption}

\newcommand{\tr}{^{\mkern-1.5mu\mathsf{T}}}

\newcommand{\ve}{\mathbb{E}}

\DeclareMathOperator{\trace}{tr}

\DeclareMathOperator{\pr}{Pr}

\addtocounter{MaxMatrixCols}{20}

\begin{document}
\title{\huge{Optimal Estimator Design and Properties Analysis for Interconnected Systems with Asymmetric Information Structure}
\thanks{This work was financially supported by Singapore National Research Foundation via Delta-NTU Corporate Lab for Cyber Physical Systems (DELTA-NTU CORP LAB-SMA-RP9 and DELTA-NTU CORP LAB-SMA-RP14).}}

\author{Yan Wang, Junlin Xiong, Zaiyue Yang, Rong Su
\thanks{Y. Wang is with the School of Mechanical Engineering and Automation, Harbin Institute of Technology Shenzhen, Shenzhen 518055, China, and is also with
the School of Electrical and Electronic Engineering, Nanyang Technological University, 50 Nanyang Avenue, Singapore. (E-mail:{\small wang.yan@hit.edu.cn})}
\thanks{
J. Xiong is with the Department of Automation, University of Science and Technology of China, Hefei 230026, China. (E-mail:{\small junlin.xiong@gmail.com})}

\thanks{Z. Yang is with Department of Mechanical and Energy Engineering, Shenzhen Key Laboratory of Biomimetic Robotics and Intelligent Systems, Guangdong Provincial Key Laboratory of Human-Augmentation and Rehabilitation Robotics in Universities, Southern University of Science and Technology, Shenzhen, China.
(E-mail:{\small yangzy3@sustech.edu.cn})}

\thanks{R. Su is with the School of Electrical and Electronic Engineering, Nanyang Technological University, 50 Nanyang Avenue, Singapore. (E-mail:{\small rsu@ntu.edu.sg})}
}

\maketitle

\IEEEpeerreviewmaketitle

\begin{abstract}
This paper studies the optimal state estimation problem for interconnected systems. Each subsystem can obtain its own measurement in real time, while, the measurements transmitted between the subsystems suffer from random delay.
The optimal estimator is analytically designed for minimizing the conditional error covariance.
The boundedness of the expected error covariance (EEC) is analyzed.
In particular, a new condition that is easy to verify is established for the boundedness of EEC. Further, the properties of EEC with respect to the delay probability are studied. We found that there exists a critical probability such that the EEC is bounded if the delay probability is below the critical probability. Also, a lower and upper bound of the critical probability is derived.
Finally, the proposed results are applied to a power system, and the effectiveness of the designed methods is illustrated by simulations.
\end{abstract}

\begin{IEEEkeywords}
Interconnected systems, expected error covariance, subsystems, random delay, optimal state estimation.
\end{IEEEkeywords}

\section{Introduction}
State estimation plays an important role in numerous applications such as
target tracking \cite{li2020distributed}, control \cite{danielson2022extremum}, and signal processing \cite{ding2022noncontact}.
With the development of the wireless network and sensor technologies, the networked state estimation
have received considerable attention during past decade.
The estimation performance is significantly affected by the
network environment.

The network attack is one of the factors having significant impact on the performance of the networked state estimation.
The remote state estimation (RSE) under denial-of-service (DoS) attacks was studied by a stochastic game framework in \cite{Ding2017}.
The nonstationary
filtering framework was designed for uncertain fuzzy Markov switching affine systems with deception attacks in \cite{cheng2022nonstationary}.
The authors of \cite{Chenbo2019} investigated the distributed dimensionality reduction fusion estimation problem for cyber-physical systems under DoS attacks.
The results of \cite{Chenbo2019} are only for a single system with multi-sensors.

The packet drops and network delays are other two major factors affecting the networked state estimation performance.
The researchers tried to understand or counteract the effects of the packet
drops/delays on the estimation performance.  The Kalman filtering with (partial) random packet drops was investigated in \cite{sinopoli2004kalman,liu2004kalman}.  The distributed Kalman filtering with multi-sensors in the presence of packet drops was studied in \cite{Jia2019}.
The results of \cite{sinopoli2004kalman,liu2004kalman,Jia2019}
are for Bernoulli packet drops model.
The Kalman filtering with Markovian packet drops/delays was studied in \cite{huang2007stability,you2011mean,han2013optimal}.
The authors of \cite{cheng2022protocol} focused on the protocol-based filtering of fuzzy Markov affine systems with uncertain packet dropouts.
In general, the  estimation problems with Markovian packet drops/delays are more complex than the ones with Bernoulli packet drops/delay.
However, it is difficult to analytically discuss the estimator properties for the Markovian packet drops/delays cases.
The literature \cite{ma2011optimal} studied the state estimation over sensor networks with mixed uncertainties of random delay, packet dropouts and missing measurements.
The state estimation problem with multiple packet losses and with the unknown  varying  delayed measurements were reported in \cite{ren2019optimal} and \cite{pohjola2008measurement}, respectively.
Only a single system with one sensor or multi-sensors is considered in \cite{sinopoli2004kalman,liu2004kalman,huang2007stability,you2011mean,Jia2019,ma2011optimal,
ren2019optimal,pohjola2008measurement}, and the extensions to interconnected systems (ISs) are rarely reported in the literature. Numerous physical systems are modeled as ISs that have attracted  lots of research attentions in the last decade \cite{wang2018decentralized,souxes2019effect,wang2019globally,wang2022optimal}.
The distributed optimal  estimation problem of IS with local information is studied in \cite{wang2021distributed}. However, the optimal estimator is not explicitly designed, and the obtained condition of the error covariance being bounded is not easy to verify \cite{wang2021distributed}.

In this paper, we focus on the optimal state estimator design for ISs with random delays. A condition in term of semidefinite programming
is established to ensure the boundedness of the EEC.  For the IS, the measurements transmitted between the subsystems suffer from random delays.
To reduce the on-line computation and save the storage space, the delayed measurements will be discarded by each subsystems.
Under the above-mentioned setup, an optimal state estimator is explicitly designed. Auxiliary equations are defined to analyze the boundedness of the expected error covariance (EEC).
In addition, the relationship between the boundedness of EEC and the delay probability is studied.
The existence of a critical probability is shown, where the EEC is bounded if the delay probability is less than the critical probability.
Also, a lower and upper bound of the critical probability is successfully derived.
Finally, the effectiveness of the proposed theories is illustrated on a power system.

\textit{\textbf{Notations:}} Let $\gamma_{a:b}$ denote the sequence $\gamma_{a},~\gamma_{a+1},\cdots,\gamma_{b}$. The probability measure is denoted by $\pr(\cdot)$.
The $2$-norm of matrix $A$ is denoted by $||A||$.  The spectral radius of matrix $A$ is denoted by $\rho(A)$. The symbol $\underline{\delta}(\cdot)$ represents the minimum eigenvalue of a matrix.
The symbols $\otimes$ and $\circ$ are the operators of Kronecker product and Hadamard product, respectively.
Let $0_{n\times m}$ denote $n\times m$ zero matrix, and $0_{n\times n}$ is abbreviated as $0_{n}$. Let $1_{n\times m}$  be a $n\times m$ matrix whose all elements are $1$, and $1_{n\times n}$ is abbreviated as $1_{n}$. The $n\times n$ unit matrix is denoted by $I_{n}$.
For a function $f(\cdot)$, define $f^{n}(\cdot)=f(f^{n-1}(\cdot))$, where $f^{1}(\cdot)=f(\cdot)$. 

\section{Problem Statement}

\begin{figure}
  \centering
  \includegraphics[width=0.3\linewidth]{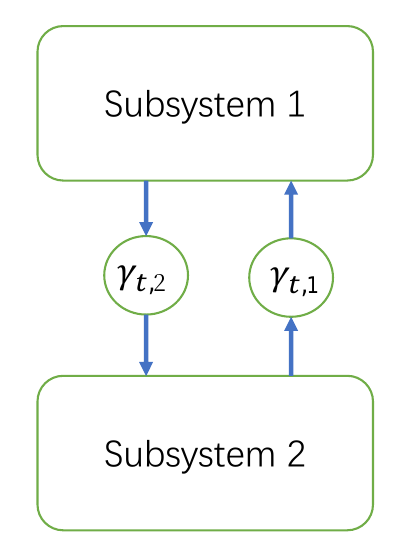}\\
  \caption{The information transmission between the subsystems: $\gamma_{t,i}=1$ means that the corresponding information transmission does not suffer delay, otherwise, $\gamma_{t,i}=0$.}\label{ff1}
\end{figure}
Consider an IS composed of two subsystems. The system dynamic is given by
\begin{subequations}
\label{zv1}
\begin{numcases}{}
\label{s1}
x_{t+1}^{1}=A^{11}x_{t}^{1}+A^{12}x_{t}^{2}+\omega_{t}^{1},\\
\label{s2}
x_{t+1}^{2}=A^{21}x_{t}^{1}+A^{22}x_{t}^{2}+\omega_{t}^{2},
 \end{numcases}
\end{subequations}
where for subsystem $i$ ($i\in\{1,2\}$), $x_{t}^{i}\in \mathbb{R}^{n_{i}}$ and $\omega_{t}^{i}\in \mathbb{R}^{n_{i}}$
are the state and process noise, respectively. The initial state $x_{0}=[(x_{0}^{1})\tr~(x_{0}^{2})\tr]\tr
$ is a random vector satisfying $\ve(x_{0})=0$ and $\ve(x_{0}\tr x_{0})=\Sigma_{0}\succ 0$. The noise
$\omega_{t}=[(\omega_{t}^{1})\tr~(\omega_{t}^{2})\tr]\tr
$
is an i.i.d. random process with $\ve(\omega_{t})=0$ and $\ve(\omega_{t}\omega_{t}\tr)=W\succ 0$.

Each subsystem employs sensors to measure its own subsystem state.  The measurement equations are given by
\begin{align}
y_{t}^{i}=C^{i}x_{t}^{i}+\upsilon_{t}^{i},~i\in\{1,2\},
\end{align}
where $\upsilon_{t}^{i}\in\mathbb{R}^{m_{i}}$, ($i\in\{1,2\}$) is the measurement noise, and
$\upsilon_{t}=[(\upsilon_{t}^{1})\tr~(\upsilon_{t}^{2})\tr]\tr$ is an i.i.d. random process satisfying $\ve(\upsilon_{t})=0$ and $\ve(\upsilon_{t}\upsilon_{t}\tr)=V\succ 0$; $C^{i}$ is the measurement matrix with a proper dimension, for $i\in\{1,2\}$. It is assumed that
$\omega_{t_{1}}$ is independent of $v_{t_{2}}$ for any $t_{1},t_{2}\geq 0$.

As Fig.~\ref{ff1} shows, subsystem $i$ will transmit the measurement $y_{t}^{i}$  to subsystem $j$ through network for $i\neq j$. The communication network between different subsystems suffers from random one step delay (one step delay or no delay).
Define the random binary variables $\gamma_{t,1}$ and $\gamma_{t,2}$ to describe the random delay. In particular,
$\gamma_{t,i}=0$ means that the measurement $y_{t}^{j}$ transmitted from subsystem $j$ to subsystem $i$ suffers from one step delay, and $\gamma_{t,i}=1$  indicates that there is no delay, where $i,j\in\{1,2\}$.
The subsystem $i$ will broadcast the value of
$\gamma_{t,i}$ to the subsystem $j$ once subsystem $i$ knows the information transmission outcomes.
Because the realization of $\gamma_{t,i}$ takes a value of either $0$ or $1$, it is easy to broadcast the value of $\gamma_{t,i}$.
Consider that the delay indicator $\gamma_{t,i}$ ($i\in\{1,2\}$) is a i.i.d. Bernoulli process with
\begin{subequations}
\label{zv3}
\begin{numcases}{}
\label{c23}
\pr(\gamma_{t,1}=1)=1-\lambda_{1},~\pr(\gamma_{t,1}=0)=\lambda_{1},\\
\label{c24}
\pr(\gamma_{t,2}=1)=1-\lambda_{2},~\pr(\gamma_{t,2}=0)=\lambda_{2},
 \end{numcases}
\end{subequations}
where $0\leq\lambda_{1},\lambda_{2}\leq 1$.
Due to the random delays,
the real time measurements available to subsystem $1$ and subsystem $2$ are $\{y_{t}^{1}, \gamma_{t,1}y_{t}^{2}\}$, and  $\{y_{t}^{2}, \gamma_{t,2}y_{t}^{1}\}$, respectively, which are referred to as \textit{asymmetric information} sets.
Using the available real time measurements, the estimator $1$ and estimator $2$ are designed as the Kalman-like filtering form:

\begin{subequations}
\label{311}
\begin{numcases}{}
\label{hx7}
\hat{x}_{t|t-1}^{1}=A^{11}\hat{x}_{t-1|t-1}^{1}+A^{12}\hat{x}_{t-1|t-1}^{2},\\
\hat{x}_{t|t}^{1}=\hat{x}_{t|t-1}^{1}+K_{t}^{11}\phi_{t}^{1}+\gamma_{t,1}K_{t}^{12}\phi_{t}^{2},
  \end{numcases}
\end{subequations}

\begin{subequations}
\label{311}
\begin{numcases}{}
\label{hx8}
\hat{x}_{t|t-1}^{2}=A^{21}\hat{x}_{t-1|t-1}^{1}+A^{22}\hat{x}_{t-1|t-1}^{2},\\
\label{ceq8}
\hat{x}_{t|t}^{2}=\hat{x}_{t|t-1}^{2}+\gamma_{t,2}K_{t}^{21}\phi_{t}^{1}+K_{t}^{12}\phi_{t}^{2},
  \end{numcases}
\end{subequations}
where $\phi_{t}^{1}=y_{t}^{1}-C^{1}\hat{x}_{t|t-1}^{1}$, $\phi_{t}^{2}=y_{t}^{2}-C^{2}\hat{x}_{t|t-1}^{2}$,
$\hat{x}_{0|0}^{1}=0_{n_{i}\times 1}$, $\hat{x}_{0|0}^{2}=0_{n_{2}\times 1}$;
$K_{t}^{ij}$, $i,j\in\{1,2\}$, are the gain matrices with proper dimensions.
Define $\hat{x}_{t|t}=[(\hat{x}_{t|t}^{1})\tr~(\hat{x}_{t|t}^{2})\tr]\tr$ and
$\hat{x}_{t|t-1}=[(\hat{x}_{t|t-1}^{1})\tr~ (\hat{x}_{t|t-1}^{1})\tr]\tr$.
The estimator of the IS is written as
\begin{subequations}
\label{zz9}
\begin{numcases}{}
\label{s11}
\hat{x}_{t|t-1}=A\hat{x}_{t-1|t-1},~~\hat{x}_{0|0}=0,\\
\label{hx14}
\hat{x}_{t|t}=\hat{x}_{t|t-1}+L_{t}(y_{t}-C\hat{x}_{t|t-1}),
  \end{numcases}
\end{subequations}
where $A=[A^{ij}]_{i,j\in\{1,2\}}$, $C=\textmd{diag}\{C^{1},~C^{2}\}$, and $L_{t}=
\begin{bmatrix}
             K_{t}^{11} & \gamma_{t,1}K_{t}^{12} \\
             \gamma_{t,2}K_{t}^{21} & K_{t}^{22} \\
           \end{bmatrix}$.
Denote $x_{t}=[( x_{t}^{1})\tr~( x_{t}^{2})\tr]\tr$.
The estimation error and prediction error are defined as
$e_{t|t}=x_{t}-\hat{x}_{t|t}$, and $e_{t|t-1}=x_{t}-\hat{x}_{t|t-1}$, respectively. The conditional error covariances are defined:
\begin{subequations}
\label{}
\begin{numcases}{}
P_{t|t}=\ve(e_{t|t}e_{t|t}\tr\big|\gamma_{t},P_{t|t-1}),\\
P_{t|t-1}=\ve(e_{t|t-1}e_{t|t-1}\tr\big|P_{t-1|t-1}),
  \end{numcases}
\end{subequations}
where $\gamma_{t}=[\gamma_{t,1}~\gamma_{t,2}]$.
Combining \eqref{s1}--\eqref{s2} and \eqref{s11}--\eqref{hx14}, one has
\begin{subequations}
\label{}
\begin{numcases}{}
e_{t|t-1}=Ae_{t-1|t-1}+\omega_{t-1},\\
e_{t|t}=e_{t|t-1}-L_{t}(Ce_{t|t-1}+\upsilon_{t}),
 \end{numcases}
\end{subequations}
and
\begin{subequations}
\label{zv9}
\begin{numcases}{}
 \label{p15}
 P_{t|t-1}=AP_{t-1|t-1}A\tr+W,~~P_{0|0}=\Sigma_{0},\\
 \label{p16}
 P_{t|t}=(I-L_{t}C)P_{t|t-1}(I-L_{t}C)\tr+L_{t}VL_{t}\tr,
 \end{numcases}
\end{subequations}
where $P_{t+1|t}$, $P_{t|t}$ are random matrices induced by the random variables $\gamma_{0:t}$.
For the IS \eqref{zv1}, we make the following definition and assumption:
\begin{definition}
A system with parameters ($A$, $C$) is detectable with $\Omega$ if there exists a $K$ such that $\rho(A-(K\circ \Omega)C)<1$.
\end{definition}
\begin{assumption}
In this paper, we assume that ($A$, $C$) is detectable with $1_{(n_{1}+n_{2})\times(m_{1}+m_{2})}$, and is undetectable with $\textmd{diag}\{1_{n_{1}\times m_{1}}, 1_{n_{2}\times m_{2}}\}$.
\end{assumption}

Under Assumption 1, $P_{t|t-1}$ is bounded if $\lambda_{1}=\lambda_{2}=0$ and is unbounded
if $\lambda_{1}=\lambda_{2}=1$, where $P_{t|t-1}$ is viewed as a nonrandom matrix if $\lambda_{1}, \lambda_{2} \in \{0, 1\}$.

\section{Optimal estimator design}
In this section, the optimal gains of the estimator are analytically derived, and the estimator realization algorithm is presented.
\begin{theorem}
\label{the1}
Consider the system \eqref{zv1}, the estimator \eqref{zz9} and the conditional error covariance dynamics \eqref{zv9}. Given $P_{t|t-1}$, $\gamma_{t,1}$, $\gamma_{t,2}$, the optimal $L_{t}$ minimizing $P_{t|t}$ is given by
\begin{align}
\label{ss12}
L_{t}&=\gamma_{t,1}\gamma_{t,2}L_{t}^{[11]}+(1-\gamma_{t,1})\gamma_{t,2}L_{t}^{[01]}\nonumber\\
&\quad+\gamma_{t,1}(1-\gamma_{t,2})L_{t}^{[10]}+(1-\gamma_{t,1})(1-\gamma_{t,2})L_{t}^{[00]},
\end{align}
where
\begin{align*}
L_{t}^{[11]}&=P_{t|t-1}C\tr(V+CP_{t|t-1}C\tr)^{-1},\\
L_{t}^{[01]}&=\begin{bmatrix}
                ( N^{3} J_{t}^{1} J_{t}^{2} U_{t}^{1}+ U_{t}^{2}) U_{t}^{3} & \begin{bmatrix}
                                                                                            0_{n_{1} \times m_{2}} \\
                                                                                            - J_{t}^{1} J_{t}^{2} \\
                                                                                          \end{bmatrix}
                 \\
               \end{bmatrix},\\
L_{t}^{[10]}&=\begin{bmatrix}
                  \begin{bmatrix}
                    - J_{t}^{3} J_{t}^{4} \\
                     0_{n_{2}\times m_{1}} \\
                   \end{bmatrix}
                   & ( N^{4} J_{t}^{3} J_{t}^{4} U_{t}^{4}+ U_{t}^{5}) U_{t}^{6}\\
                \end{bmatrix},\\
L_{t}^{[00]}&=\begin{bmatrix}
                  { N^{4}}\tr U_{t}^{2} U_{t}^{3} & 0_{n_{1}\times m_{2}} \\
                  0_{n_{2}\times m_{1}} & { N^{3}}\tr U_{t}^{5} U_{t}^{6} \\
                \end{bmatrix},\\
 N^{1}&=[I_{m_{1}}~0_{m_{1}\times m_{2}}],~  N^{2}=[0_{m_{2}\times m_{1}}~I_{m_{2}}],\\
 N^{3}&=\begin{bmatrix}
           0_{n_{1}\times n_{2}} \\
           I_{n_{2}} \\
         \end{bmatrix},~
 N^{4}=\begin{bmatrix}
              I_{n_{1}} \\
              0_{n_{2}\times n_{1}} \\
            \end{bmatrix},\\
 J_{t}^{1}&={ N^{3}}\tr \Big( U_{t}^{2} U_{t}^{3} U_{t}^{4} U_{t}^{6}- U_{t}^{5} U_{t}^{6}\Big),\\
 J_{t}^{2}&=\Big(I- U_{t}^{1} U_{t}^{3} U_{t}^{4} U_{t}^{6}\Big)^{-1},~J_{t}^{4}=(I- U_{t}^{4} U_{t}^{6} U_{t}^{1} U_{t}^{3})^{-1},\\
 J_{t}^{3}&={ N^{4}}\tr( U_{t}^{5} U_{t}^{6} U_{t}^{1} U_{t}^{3}- U_{t}^{2} U_{t}^{3}),\\
 U_{t}^{1}&= N^{2}V{ N^{1}}\tr+ N^{2}CP_{t|t-1}C\tr{ N^{1}}\tr,\\
 U_{t}^{2}&=P_{t|t-1}C\tr{ N^{1}}\tr,~U_{t}^{5}=P_{t|t-1}C\tr{ N^{2}}\tr,\\
 U_{t}^{3}&=\Big({ N^{1}}V{ N^{1}}\tr+{ N^{1}}CP_{t|t-1}C\tr{ N^{1}}\tr\Big)^{-1},\\
 U_{t}^{4}&= N^{1}V{ N^{2}}\tr+ N^{1}CP_{t|t-1}C\tr{ N^{2}}\tr,\\
 U_{t}^{6}&=\Big({ N^{2}}V{ N^{2}}\tr+{ N^{2}}CP_{t|t-1}C\tr{ N^{2}}\tr\Big)^{-1}.
\end{align*}
\end{theorem}
\begin{IEEEproof}
See appendix.
\end{IEEEproof}

The optimal state estimator of IS \eqref{zv1} with random delay is realized by Algorithm 1. Before presenting Algorithm 1, denote
$\bar{ N}^{1}=(N^{4})\tr$, $\bar{ N}^{2}=(N^{1})\tr$, $\bar{ N}^{3}=(N^{2})\tr$, $\bar{ N}^{4}=(N^{3})\tr$.
\begin{algorithm}
\caption{Sub-estimator realization}
\begin{algorithmic}
\STATE 1.~Obtain $y_{t}^{i}$ and transmit $y_{t}^{i}$ to subsystem $j$, where $i,j\in\{1,2\}$ and $i\neq j$.
\STATE 2.~Check whether $y_{t}^{j}$ is obtained in real time; broadcast the value of $\gamma_{t,i}$ to subsystem $j$, and determine the value of $\gamma_{t,j}$.
\STATE 3.~Compute $L_{t}^{[\gamma_{t,1}\gamma_{t,2}]}$ based on Theorem 1 using $\gamma_{t,1}$, $\gamma_{t,2}$, and $P_{t|t-1}$.
\STATE 4.~Compute $\hat{x}_{t|t-1}^{1}$ and $\hat{x}_{t|t-1}^{2}$ by \eqref{hx7}, \eqref{hx8} using $\hat{x}_{t-1|t-1}^{1}$ and $\hat{x}_{t-1|t-1}^{2}$.\\
\begin{spacing}{0.8}
\dotfill
\end{spacing}
\textbf{for subsystem 1:}
\STATE 5.~Compute $
 \hat{x}_{t|t}^{1}=\hat{x}_{t|t-1}^{1}+\bar{ N}^{1}L_{t}^{[\gamma_{t,1}\gamma_{t,2}]}\bar{ N}^{2}(y_{t}^{1}-C^{1}\hat{x}_{t|t-1}^{1})
$.

\IF{$\gamma_{t,1}=1$}
\STATE  6.~Update
$
 \hat{x}_{t|t}^{1}=\hat{x}_{t|t}^{1}+\bar{ N}^{1}L_{t}^{[\gamma_{t,1}\gamma_{t,2}]}\bar{ N}^{3}(y_{t}^{2}-C^{2}\hat{x}_{t|t-1}^{2})
$.
\ENDIF
\begin{spacing}{0.8}
\dotfill
\end{spacing}
\begin{spacing}{0.8}
\dotfill
\end{spacing}
\textbf{for subsystem 2:}
\STATE 5.~Compute $
 \hat{x}_{t|t}^{2}=\hat{x}_{t|t-1}^{2}+\bar{ N}^{4}L_{t}^{[\gamma_{t,1}\gamma_{t,2}]}\bar{ N}^{3}(y_{t}^{2}-C^{2}\hat{x}_{t|t-1}^{2})
$.

\IF{$\gamma_{t,2}=1$}
\STATE 6.~Update
$
 \hat{x}_{t|t}^{2}=\hat{x}_{t|t}^{2}+\bar{ N}^{4}L_{t}^{[\gamma_{t,1}\gamma_{t,2}]}\bar{ N}^{2}(y_{t}^{1}-C^{1}\hat{x}_{t|t-1}^{1})
$.
\ENDIF
\begin{spacing}{0.5}
\dotfill
\end{spacing}
\STATE 7.~Transmit $\hat{x}_{t|t}^{i}$ to subsystem $j$.
\STATE 8.~Compute $P_{t+1|t}$ based on \eqref{p15}--\eqref{p16}.
\STATE 9.~Let $t=t+1$ and return to first step.
\end{algorithmic}
\end{algorithm}

\begin{remark}
\label{cr3}
Compared to \cite{sinopoli2004kalman,liu2004kalman,huang2007stability,you2011mean}, the derivation of the optimal estimator gain $L_{t}$ in this paper is more challenging, because if $\gamma_{t,1}\gamma_{t,2}\neq1$, the corresponding gains of the estimator must satisfy a certain sparse structure constraint. While, the optimal estimator gain in \cite{sinopoli2004kalman,liu2004kalman,huang2007stability,you2011mean} is directly obtained from standard Kalman filtering formulas. Actually, for any cases of the data reception, the optimal estimator gain in \cite{sinopoli2004kalman,liu2004kalman,huang2007stability,you2011mean} is either a zero matrix or the standard Kalman filtering gain with different measurement matrices.
\end{remark}

\section{Boundedness analysis of expected error covariance}

In this section, we analyse the boundedness of the EEC through auxiliary matrix functions.
For simplicity, hereinafter, we denote $P_{t}= P_{t|t-1}$. It follows from \eqref{zv9} that
\begin{subequations}
\label{zv11}
\begin{numcases}{}
\label{ep}
P_{t+1}=A_{L,t}P_{t}A_{L,t}\tr+AL_{t}V(AL_{t})\tr+W_{t},\\
P_{1}=A\Sigma_{0}A\tr+W,
 \end{numcases}
\end{subequations}
where $A_{L,t}=A-AL_{t}C$.
In the following, we focus on studying the properties of $\ve(P_{t})$.  According to the definition of $P_{t}$, $\ve(P_{t})$ depends on the distributions of the random variables $\gamma_{t,1}$, $\gamma_{t,2}$.
Define the following auxiliary matrix functions
\begin{align}
\label{bb1}
a(X,Y)&=(A-AXC)Y(A-AXC)\tr,\\
\label{c20}
b(X,Y)&=a(X,Y)+AXVX\tr A\tr+W,\\
\label{cc21}
c(X)&=(A-AXC)\tr(A-AXC),\\
\label{zz15}
 d(X)&=(A-AXC)\otimes (A-AXC).
\end{align}
Recall Theorem 1.
Note that $L_{t}^{\gamma}$, $\gamma\in\{[00],[01],[10],[11]\}$ depends on $P_{t}$, and thus we denote $L_{t}^{\gamma}$ by $L^{\gamma}[P_{t}]$, where we take $P_{t}$ as a variable. Consider the probability distribution of $\gamma_{t}=[\gamma_{t,1}~\gamma_{t,2}]$, i.e. \eqref{zv3}, we define
\begin{align}
\label{g37}
&  f(S^{1},S^{2},S^{3},S^{4},Y)=\lambda_{1}\lambda_{2}b(S^{1},Y)+\lambda_{1}(1-\lambda_{2})b(S^{2},Y)\nonumber\\
&+(1-\lambda_{1})\lambda_{2}b(S^{3},Y)+(1-\lambda_{1})(1-\lambda_{2})b(S^{4},Y),\\
\label{g38}
&g_{\lambda_{1}\lambda_{2}}(Y)=  f( L^{[00]}[Y], L^{[01]}[Y], L^{[10]}[Y], L^{[11]}[Y],Y).
\end{align}
Some useful properties of the the matrix equation \eqref{g38} are proposed below (Proposition 1 and Lemmas \ref{elem1}--\ref{elem2}) and will be used to establish the boundedness condition of EEC.
\begin{proposition}
\label{lem1}
Consider \eqref{zv11} and \eqref{g38}.
The following equations hold:
\begin{align}
\label{g1}
\ve(P_{t+1}|P_{t})&=g_{\lambda_{1}\lambda_{2}}(P_{t}),~
\ve(P_{t+1})=\ve(g_{\lambda_{1}\lambda_{2}}(P_{t})),\\
\label{g3}
\ve(P_{t+1})&=\ve(g_{\lambda_{1}\lambda_{2}}(P_{t}))\preceq g_{\lambda_{1}\lambda_{2}}(\ve(P_{t})).
\end{align}
\end{proposition}
\begin{IEEEproof}
See appendix.
\end{IEEEproof}

\begin{lemma}
\label{elem1}
Define the matrix  sequence $Y_{0},~Y_{1},~Y_{2},\cdots$ generated by
$
Y_{t+1}=g_{\lambda_{1}\lambda_{2}}(Y_{t}),~Y_{0}=P_{0}$. One has $\ve(P_{t})\preceq Y_{t}$.
\end{lemma}
\begin{IEEEproof}
This result directly follows from \eqref{g3}.
\end{IEEEproof}

\begin{lemma}
\label{elem2}
Consider the sequence $Y_{0},~Y_{1},~Y_{2},\cdots$ defined in Lemma \ref{elem1}. The following result holds:
$\lim\limits_{t\rightarrow +\infty}Y_{t}$ is bounded if and only if there exists $X^{1},\cdots,X^{7}$ such that $\rho\Big(  h(X^{1},\cdots,X^{7})\Big)<1$,
where
\begin{align}
\label{e41}
&h(X^{1},\cdots,X^{7})
=\lambda_{1}\lambda_{2} d( N^{4}X^{1} N^{1}+ N^{3}X^{2} N^{2})\nonumber\\
&\quad+\lambda_{1}(1-\lambda_{2}) d(X^{3} N^{1}+ N^{3}X^{4} N^{2})\nonumber\\
&\quad+(1-\lambda_{1})\lambda_{2}  d( N^{4}X^{4} N^{1}+X^{6} N^{2})\nonumber\\
&\quad+(1-\lambda_{1})(1-\lambda_{2}) d(X^{7}).
\end{align}
\end{lemma}
\begin{IEEEproof}
See appendix.
\end{IEEEproof}

In the following, we derive a sufficient condition of $\rho\big(  h(X^{1},\cdots,X^{7})\big)<1$ with the form of semidefinite programming (SDP).

\begin{theorem}
\label{zz2}
Consider \eqref{zv11} with \eqref{zv3}.
Given $\lambda_{1},~\lambda_{2}$.
If there exists $r_{1}, r_{2}, r_{3}, r_{4}$ satisfying $r_{1}\lambda_{1}\lambda_{2}+r_{2}\lambda_{1}(1-\lambda_{2})+r_{3}(1-\lambda_{1})\lambda_{2}+r_{4}(1-\lambda_{1})(1-\lambda_{2})< 1$, where
\begin{subequations}
\label{}
\begin{numcases}{}
\label{e33}
r_{1}=\textmd{arg}\min_{r,X,\tilde{X}}\Big\{r\geq0:~ p_{1}(r,X,\tilde{X})\succeq 0\Big\},\\
r_{2}=\textmd{arg}\min_{r,X,\tilde{X}}\Big\{r\geq0:~ p_{2}(r,X,\tilde{X})\succeq 0\Big\},\\
\label{zz33}
r_{3}=\textmd{arg}\min_{r,X,\tilde{X}}\Big\{r\geq0:~ p_{3}(r,X,\tilde{X})\succeq 0\Big\},\\
\label{e36}
r_{4}=\textmd{arg}\min_{r,X}\Big\{r\geq0:~ p(r,X)\succeq 0\Big\},
 \end{numcases}
\end{subequations}
\begin{subequations}
\label{}
\begin{numcases}{}
\label{e37}
 p(r,X)=\begin{bmatrix}
                          \sqrt{r} I & A-AXC  \\
                         (A-AXC)\tr  & \sqrt{r}I \\
                        \end{bmatrix},\\
\label{e38}
 p_{1}(r,X,\tilde{X})= p(r, N^{4}X N^{1}+ N^{3}\tilde{X} N^{2}),\\
 p_{2}(r,X,\tilde{X})= p(r,X N^{1}+ N^{3}\tilde{X} N^{2}),\\
 p_{3}(r,X,\tilde{X})= p(r, N^{4}X N^{1}+\tilde{X} N^{2}),
 \end{numcases}
\end{subequations}
then $\lim\limits_{t\rightarrow +\infty}\ve(P_{t})$  is bounded.
\end{theorem}
\begin{IEEEproof}
See appendix.
\end{IEEEproof}
\begin{remark}
Note that $ p(r,X)\succeq 0$, $ p_{1}(r,X,\tilde{X})\succeq 0$, $ p_{2}(r,X,\tilde{X})\succeq 0$, and $ p_{3}(r,X,\tilde{X})\succeq 0$ are LMIs. Problems \eqref{e33}--\eqref{e36} are SDP problems which can be effectively solved by MATLAB toolbox. Thus, the condition established in Theorem 2 can be effectively verified.
\end{remark}


\begin{corollary}
\label{coro}
 Consider $r_{1}$, $r_{2}$, $r_{3}$ and $r_{4}$ defined in Theorem \ref{zz2}. It holds that  $r_{4}\leq \min(r_{2},r_{3})$, $r_{1}\geq\max(r_{2},r_{3})$. In addition, if $r_{1}=r_{4}$ and there exists $r_{i}< 1$, for any $i\in\{1,2,3,4\}$, then $\lim\limits_{t\rightarrow +\infty}\ve(P_{t})$ is  bounded for any $\lambda_{1}$, $\lambda_{2}\in[0~ 1]$.
\end{corollary}
\begin{IEEEproof}
See appendix.
\end{IEEEproof}

\begin{corollary}
\label{coro2}
Consider $r_{1}$, $r_{2}$, $r_{3}$ and $r_{4}$ defined in Theorem \ref{zz2}. Assume that the system parameters $(A~C)$ satisfy Assumption 1. If there exists a $X$ satisfying $c(X)\prec I$, then $r_{4}<r_{1}$, where $c(\cdot)$ is defined in \eqref{cc21}.
\end{corollary}
\begin{IEEEproof}
See appendix.
\end{IEEEproof}

Note that $\ve(P_{t})$ depends on $\lambda_{1},~\lambda_{2}$.
It is known that if the delay occurs with a bigger probability, then the estimation performance gets worse. This means that $\ve(P_{t})$ is monotone increasing with respect to $\lambda_{1},~\lambda_{2}$.
Additionally, if $\lambda_{1},~\lambda_{2}$ are big enough, $\ve(P_{t})$ may
became unbounded when the time goes to infinity. The above discussions are concluded as follows.

\begin{lemma}
\label{lee6}
Consider \eqref{zv11}.
For a fixed $\lambda_{1}$, if $\lim\limits_{t\rightarrow\infty}\ve(P_{t})$ is unbounded for $\lambda_{2}=1$, but bounded  for $\lambda_{2}=0$, then there exists a critical probability $\lambda_{2,c}$ such that
$\lim\limits_{t\rightarrow\infty}\ve(P_{t})$ is bounded for $\lambda_{2}\leq\lambda_{2,c}$, and unbounded for
$\lambda_{2}>\lambda_{2,c}$.
Also, for a fixed $\lambda_{2}$, we have the similar results to $\lambda_{1}$.
\end{lemma}
\begin{IEEEproof}
See appendix.
\end{IEEEproof}

Since the accurate critical probability of delay is not easy to obtain,  the lower and upper bounds of the critical probability may be useful.
A computation method is provided in the following theorem to compute the lower and upper bounds of the critical probability.
For ease of notations, we first define
\begin{align*}
&q(X^{1},X^{2})=c( N^{4}X^{1} N^{1}+ N^{3}X^{2} N^{2}).
\end{align*}

\begin{theorem}
\label{co1}
Consider \eqref{zv11}.
For a fixed $\lambda_{1}$, the critical probability $\lambda_{2,c}$ satisfies that if $\lambda_{2}\leq\lambda_{2,c}$, $\lim\limits_{t\rightarrow\infty}\ve(P_{t})$ is bounded, otherwise, $\lim\limits_{t\rightarrow\infty}\ve(P_{t})$ is unbounded. A lower and upper bound of $\lambda_{2,c}$ is obtained, i.e.
$\underline{\lambda}_{2,c}\leq\lambda_{2,c}\leq\bar{\lambda}_{2,c}$, where
\begin{align}
\label{ee4}
\bar{\lambda}_{2,c}
&=\left\{
    \begin{array}{ll}
      1, & \hbox{if $r_{1}=r_{4}< 1$;}\\
\min\Big\{\frac{1}{\alpha\lambda_{1}},1\Big\},, & \hbox{otherwise;} \\
    \end{array}
  \right.\\
\label{ee5}
\underline{\lambda}_{2,c}&=\left\{
                               \begin{array}{ll}
                                \frac{1-r_{2}\lambda_{1}-r_{4}(1-\lambda_{1})}{(r_{1}-r_{2})\lambda_{1}+(r_{3}-r_{4})(1-\lambda_{1})}, & \hbox{if $r_{1}\neq r_{4}$;} \\
                                1 , & \hbox{if $r_{1}=r_{4}< 1$;} \\
                                0 , & \hbox{if $r_{1}=r_{4}\geq 1$.}
                               \end{array}
                             \right.
\end{align}
where
$\alpha=\underline{\delta}\Big(\min\limits_{X^{1},X^{2}}q(X^{1},X^{2})\Big)$.
Similar upper and lower bounds for $\lambda_{1}^{c}$ can be obtained when we fix $\lambda_{2}$.
\end{theorem}
\begin{IEEEproof}
See appendix.
\end{IEEEproof}
Theorem \ref{co1} provides a method to compute the upper and lower bounds of the critical probability of delay.

\begin{remark}[Extension to large-scale ISs]
Consider a large-scale IS composed of $n$ subsystems.
We can use a random graph $\mathcal{G}(\mathcal{V},\mathcal{E},p)$ to describe the communication between the subsystems, where $\mathcal{V}$ is the nodes set; $\mathcal{E}$ is the possible edges set; $p=[p_{1},\ldots,p_{|\mathcal{E}|}]$. At time $t$, the $i$-th possible edge in $\mathcal{E}$ occurs in the graph with probability $0 <p_{i}<1$.
The edge $(i,j)$ occurring in the graph implies that the subsystem $i$ can receive the information of subsystem $j$ in real time.
The  continuity of $\ve(P_{t})$ (w.r.t. $p$) indicates that there should exist a critical plane denoted by $p_{c}$, such that
the boundedness of $\ve(P_{t})$ depends on which side of the plane $p_{c}$ the point $p$ is on.
However, it is not easy to determine the upper and lower bounds of $p_{c}$.
\end{remark}

\section{Applications to Power Systems}
In this section, a power system is used as an example to demonstrate the effectiveness of the proposed state estimation methods.

\begin{figure}[!h]
  \centering
  \includegraphics[width=0.7\linewidth]{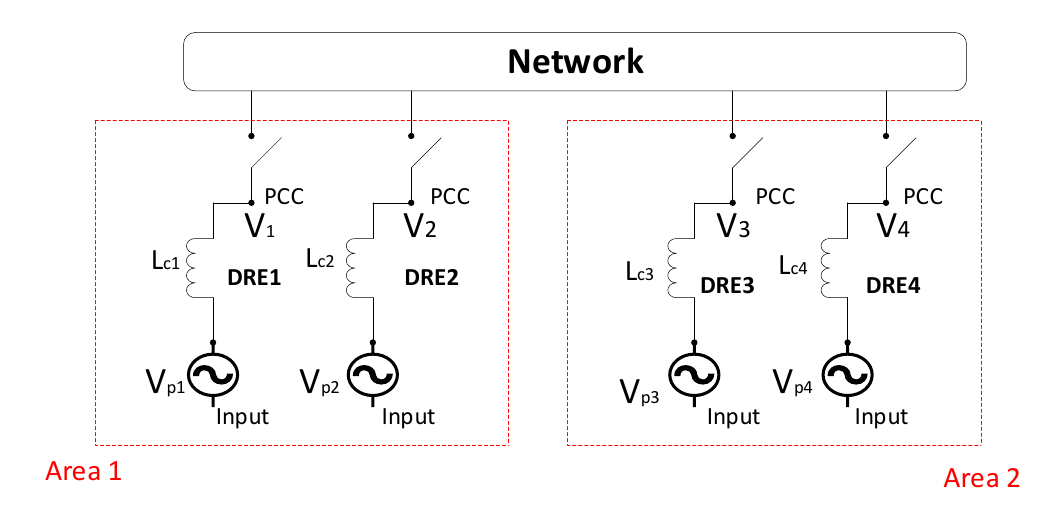}\\
  \caption{The model of distributed energy resources (DERs) connecting to the
power network \cite{Routing2012}. DER 1 and DER 2 are in area 1; DER 3 and DER 4 are in area 2.}\label{DEr}.
\end{figure}

We study the state estimation problem for the distributed energy resources (DERs) connected to the
power network that is shown in Fig.~\ref{DEr}. In this example, the power network is chosen to be the IEEE 4-bus distribution network (see Fig.~3 of \cite{Routing2012}).
As Fig.~\ref{DEr} depicts, four DERs are integrated into the main power network at the point common coupling (PCC). The voltages of the PCC are $v_{s}=[v_{1}~v_{2}~v_{3}~v_{4}]\tr$,
where $v_{i}$ is the voltage of the $i$th PCC, and $i\in\{1,2,3,4\}$. Each DER is a voltage source at each bus. The input voltage of the voltage sources is denoted by
$v_{p}=[v_{p1}~v_{p2}~v_{p3}~v_{p4}]\tr$. To maintain the proper operation of DERs, it is required to keep the PCC voltages $v_{s}$ at reference values $v_{ref}$.
The PCC voltage deviation $x_{t}=v_{s}-v_{ref}$ is chosen to be the system state. The DER control effort deviation $u_{t}=v_{p}-v_{pref}$ is the control input, where $v_{pref}$ is the reference of the control effort. Following the modeling process provided in \cite{Routing2012,Rana2017}, the dynamic of $x_{t}$ is of the following form:
\begin{align}
\label{ex}
\dot{x}_{t}=Ax_{t}+Bu_{t}+n_{t},
\end{align}
where $n_{t}$ is a zero-mean Gaussian white noise whose covariance is $Q$.
The system \eqref{ex} is discretized as:
\begin{align}
\label{dx}
x_{t+1}=A_{d}x_{t}+B_{d}u_{t}+\omega_{t},
\end{align}
where $A_{d}=I+AT_{s}$, $B_{d}=T_{s}B$, and $\omega_{t}$ is the i.i.d. Gaussian process with zero-mean and covariance $W=T_{s}Q$.
Assume the initial state $x_{0}$ is a zero mean Gaussian variable with identity covariance.
Here, we assume $W=I_{4}$, and set the sampling time $T_{s}=0.05s$.
In this example, $\rho(A_{d})>1$, which means that the system \eqref{dx} under open-loop pattern (there is no control input) is unstable. Here, the controller is of the state feedback form: $u_{t}=Fx_{t}$, where $F$ is the gain matrix with proper dimension. The closed-loop system is given by
\begin{align}
\label{ac32}
x_{t+1}=A_{c}x_{t}+ \omega_{t},
\end{align}
where $A_{c}=A_{d}+B_{d}F$.
As Fig.~\ref{DEr} shows, DERs 1--2 are in area 1 and DERs 3--4 are in area 2. Based on different areas, the system \eqref{ac32} can be written as an IS composed of two subsystems:
\begin{align}
\label{x25}
x_{t+1}^{i}=A_{c}^{ii}x_{t}^{i}+A_{c}^{ij}x_{t}^{j}+ \omega_{t}^{i},~i,j\in\{1,2\},~ i\neq j,
\end{align}
where $x_{t}^{i}$ is the component of the system state corresponding to area $i$; and $A_{c}=[A_{c}^{ij}]_{i,j\in\{1,2\}}$,
$x_{t}^{1}=[I_{2}~ 0_{2}]x_{t}$,  $x_{t}^{2}=[0_{2}~ I_{2}]x_{t}$,
$\omega_{t}^{1}=[I_{2}~ 0_{2}]\omega_{t}$,  $\omega_{t}^{2}=[0_{2}~ I_{2}]\omega_{t}$.
To monitor the working status of the power system \eqref{x25}, in each area, the sensors are employed to measure the system state.
The measurement equations of area 1 and area 2 are
\begin{align}
y_{t}^{i}=C^{i}x_{t}^{i}+ \upsilon_{t}^{i},~i\in\{1,2\},
\end{align}
where $ \upsilon_{t}=[ \upsilon_{t}^{1}~ \upsilon_{t}^{2}]\tr$ is measurement noise that is the zero-mean Gaussian independent processes with $\ve( \upsilon_{t} \upsilon_{t}\tr)=I_{3}$, and $C^{1}=[I_{2}~0_{2}]$, $C^{2}=[0~0~1~1]$.
In area $i$, the estimator $i$ is installed to estimate the state $x_{t}^{i}$. The
measurement $y_{t}^{1}$ is transmitted to estimators 1--2 through network, so does $y_{t}^{2}$.
It is assumed that
the data $y_{t}^{i}$ transmitted to estimator $j$ suffers from random delay (one step delay or no delay), where $i\neq j$ and  $i,j\in\{1,2\}$.
Assume that the  delay indicator $\gamma_{t,i}$ is a Bernoulli stochastic process with $\pr(\gamma_{t,i}=0)=\lambda_{i}$, where $i\in\{1,2\}$.

\begin{figure}[!h]
  \centering
  \includegraphics[width=0.8\linewidth]{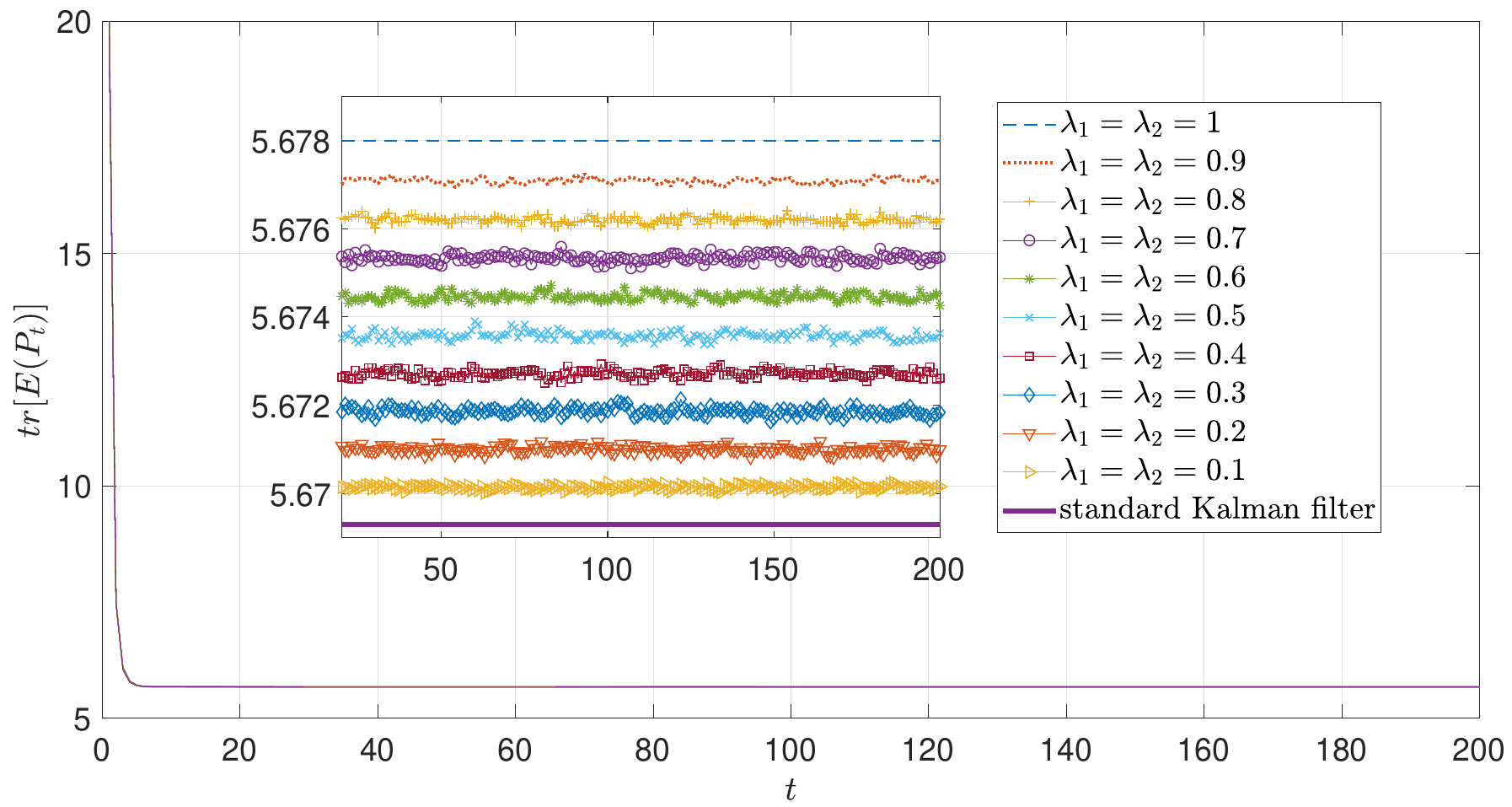}\\
  \caption{The value of $\trace[\ve(P_{t})]$ under different $\lambda_{1}$, $\lambda_{2}$.}\label{fig5}
\end{figure}

\textbf{\textit{Case 1 (the stable closed-loop system):}}
The controller gain is chosen such that
$\rho(A_{c})<1$, which means that
the system \eqref{x25} with the chosen $F$ is stable.
Then, our proposed method is applied for the state estimation of the system \eqref{x25}. The estimation performance is evaluated by the trace of the
expected error covariance $\ve(P_{t})$. The trace of $\ve(P_{t})$ with respect to different delay probabilities ($\lambda_{1}, \lambda_{2}$) is approximately computed by averaging 1000 Monte-Carlo simulations, see Fig.~\ref{fig5}. From Fig.~\ref{fig5}, we can see that (i) the estimation performance gets worse under larger delay probabilities; (ii) the expected error covariance $\ve(P_{t})$ is bounded even though $\lambda_{1}=\lambda_{2}=1$ (Note that $\lambda_{1}=\lambda_{2}=1$ means that the delay always happens); (iii)
under different delay probabilities, the expected error covariance $\ve(P_{t})$ is within the same order of magnitude as the error covariance of standard Kalman filtering.
The simulations illustrate that our estimators can well monitor the working state of the power system \eqref{x25}, even though the data transmitted between different areas suffer from random delays.

\begin{figure}[!h]
  \centering
  \includegraphics[width=0.8\linewidth]{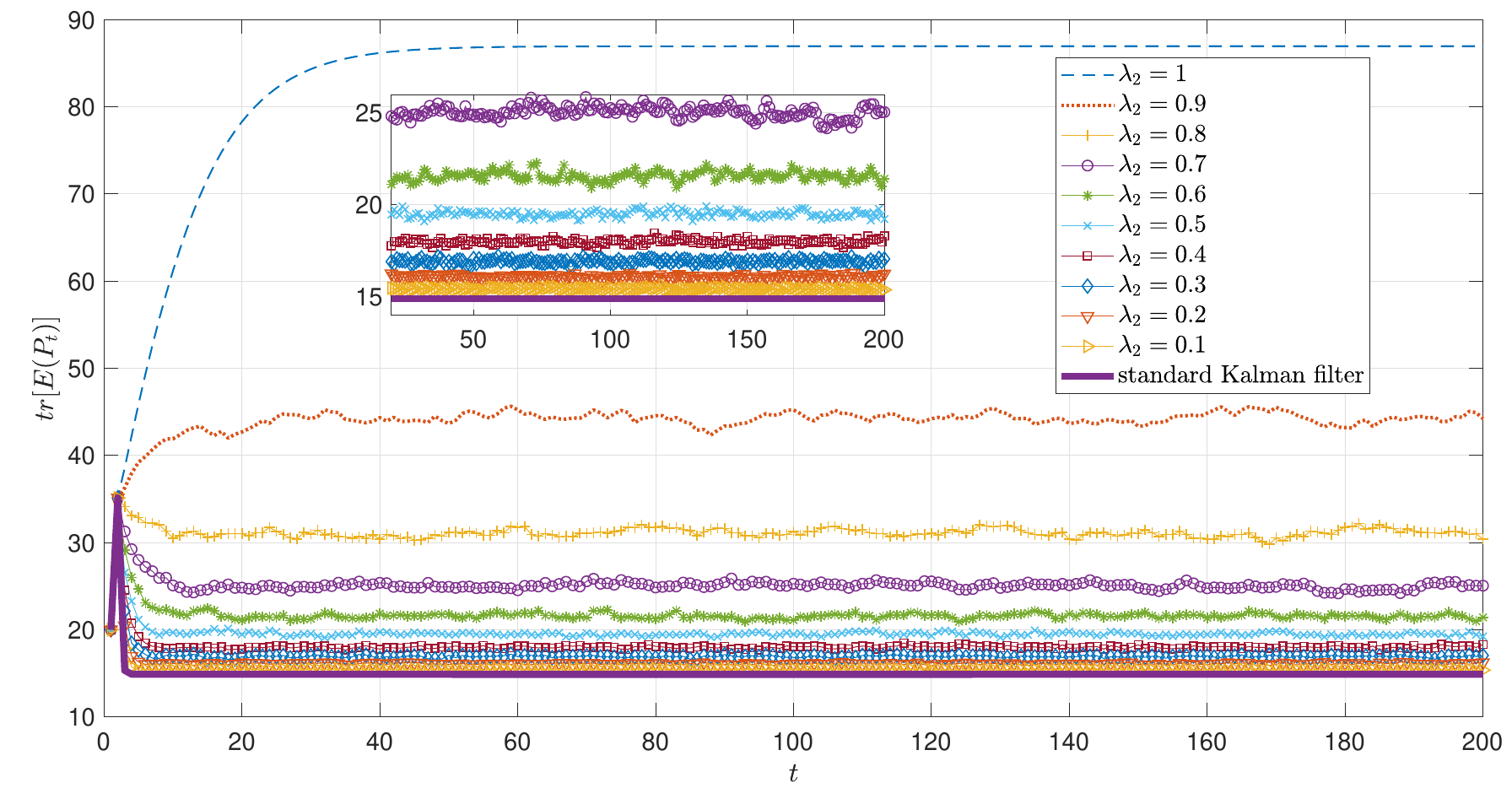}\\
  \caption{$\lambda_{1}=1$, the value of $\trace[\ve(P_{t})]$ under different $\lambda_{2}$}\label{m2}
\end{figure}

\begin{figure}[!h]
  \centering
  \includegraphics[width=0.8\linewidth]{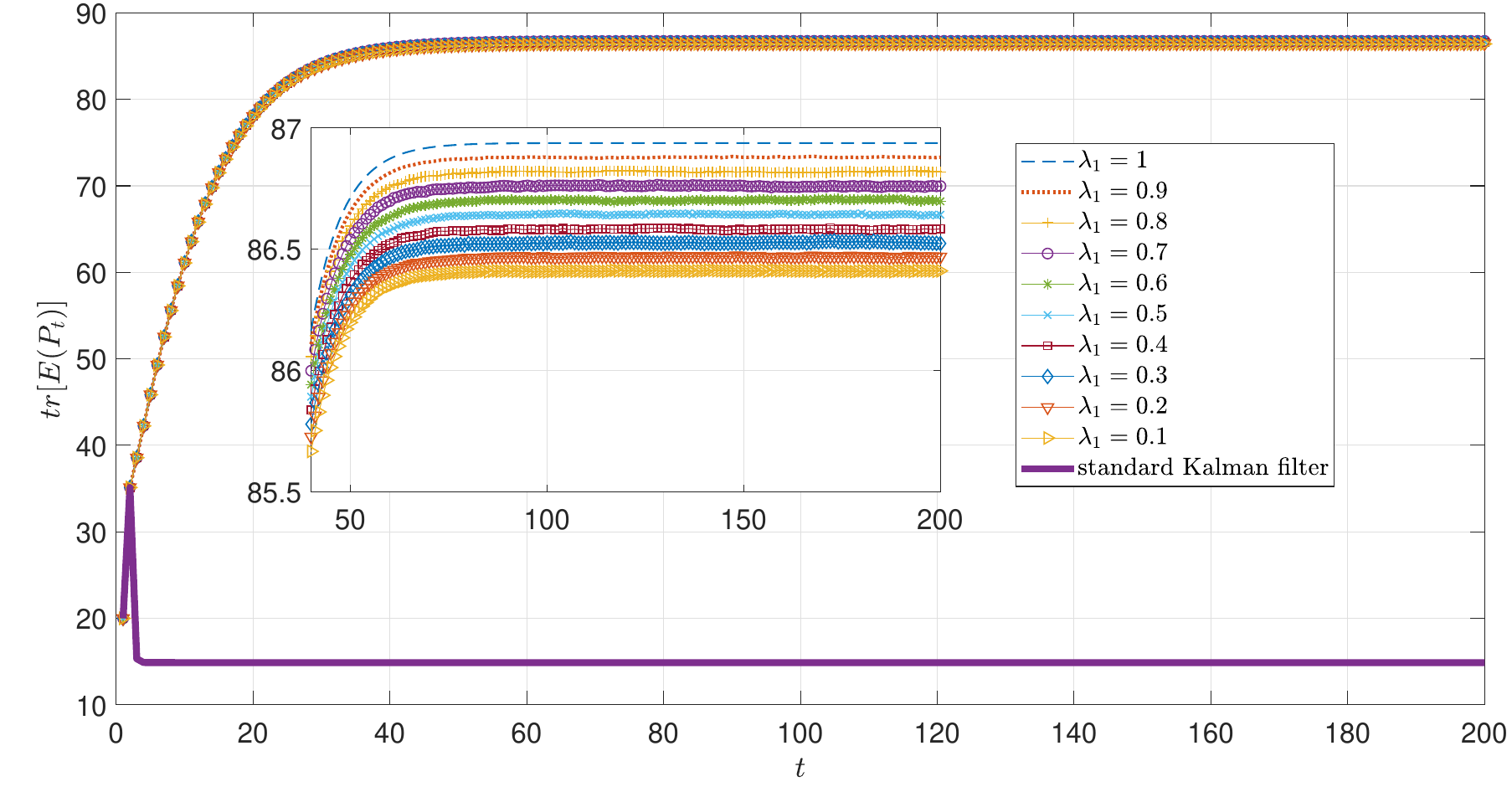}\\
  \caption{$\lambda_{2}=1$, the value of $\trace[\ve(P_{t})]$ under different $\lambda_{1}$}\label{m3}
\end{figure}

\textbf{\textit{Case 2 (the unstable closed-loop system):}} It is known that to estimate the states of unstable systems is more challenging than to estimate the states of stable systems. Hence, to further illustrate the effectiveness of our estimators, we also consider the unstable system case. The controller gain is chosen such that
$\rho(A_{c})>1$.
Hence, the system \eqref{x25} with the chosen controller gain is unstable. In this case, the values of $\trace[\ve(P_{t})]$ with different $(\lambda_{1}, \lambda_{2})$ are presented in Figs.~\ref{m2}--\ref{m3}.

Figs.~\ref{m2}--\ref{m3} show that the expected error covariance $\ve(P_{t})$ (with any delay probabilities) is within the same order of magnitude as the error covariance of standard Kalman filtering. This implies that even when the power systems \eqref{x25} become unstable, the working states are also well monitored by our estimator under random delay. In addition, comparing Fig.~\ref{m2} and Fig.~\ref{m3}, we find that for the power system with the parameters given in this case,
the estimation performance is more sensitive to $\lambda_{2}$ than to $\lambda_{1}$.

\section{Conclusion}
This paper studied the optimal estimator design problem for IS with random delay. Due to the random delay occurring among the subsystems, the information available to different subsystem may be different. This type of IS is called IS with asymmetric information structure. An optimal estimator has been analytically designed. The estimator realization algorithm for each subsystem was developed.
In addition, some useful properties of the estimation performance were obtained.
Finally, the proposed estimator was applied to a power system.
The simulations showed that the designed estimator is effective and is of good performance.

In this work, we assume the delay indicator obeying Bernoulli process. In the further work, we may extend this work to the Markov chain delay indicator case.

\bibliographystyle{IEEEtran}
\bibliography{bibfile}

\newpage

\section{Appendix}
\subsection{The proof of Theorem \ref{the1}}
\label{app22a}
The matrices $L_{t}^{[11]}$, $ U_{t}^{3}$, $ U_{t}^{6}$, $ J_{t}^{2}$, $ J_{t}^{4}$ are defined by the ``inverse'' operator.
Firstly, we show that these matrices are well-defined. It follows from $V+CP_{t|t-1}C\tr\succeq V\succ0$ that $(V+CP_{t|t-1}C\tr)^{-1}$ is well-defined. Thus, $L_{t}^{[11]}$ is well-defined. Since both $ N^{1}$ and $ N^{2}$
are row full rank, ${ N^{1}}V{ N^{1}}\tr+{ N^{1}}CP_{t|t-1}C\tr{ N^{1}}\tr\succeq{ N^{1}}V{ N^{1}}\tr\succ0$,
${N^{2}}V{N^{2}}\tr+{N^{2}}CP_{t|t-1}C\tr{N^{2}}\tr\succeq{ N^{2}}V{ N^{2}}\tr\succ0$. Thus, $ U_{t}^{3}$ and $ U_{t}^{6}$ are well-defined. Based on the Schur complement decomposition, it follows from
\begin{align*}
V+CP_{t|t-1}C\tr=\begin{bmatrix}
                   ( U_{t}^{3})^{-1} &  U_{t}^{4} \\
                    U_{t}^{1} & ( U_{t}^{6})^{-1} \\
                 \end{bmatrix}\succ0
\end{align*}
that $ U_{t}^{1} U_{t}^{3} U_{t}^{4}\prec( U_{t}^{6})^{-1}$, and $ U_{t}^{4} U_{t}^{6} U_{t}^{1}\prec( U_{t}^{3})^{-1}$. As a result,
$I- U_{t}^{1} U_{t}^{3} U_{t}^{4} U_{t}^{6}\succ0$, $I- U_{t}^{4} U_{t}^{6} U_{t}^{1} U_{t}^{3}\succ0$. This shows that $ J_{t}^{2}$ and $ J_{t}^{4}$ are well-defined.

Now, we start to prove that the optimal $L_{t}$ is of the form \eqref{ss12}.
For ease of notation, define
$
 R_{t}^{1}=\begin{bmatrix}
                       K_{t}^{11} \\
                       K_{t}^{21} \\
                     \end{bmatrix}$,
$ R_{t}^{2}=\begin{bmatrix}
                       K_{t}^{12} \\
                       K_{t}^{22}\\
                     \end{bmatrix}$.

(\textbf{Case 1}) If $\gamma_{t,1}=1$, $\gamma_{t,2}=1$, then the optimal $L_{t}=L_{t}^{[11]}$ is the gain of the standard Kalman filtering \cite{aastrom2012introduction}.

(\textbf{Case 2}) If $\gamma_{t,1}=0$, $\gamma_{t,2}=1$, then $L_{t}$ has the form $L_{t}= R_{t}^{1} N^{1}+ N^{3}K_{t}^{22} N^{2}$. Inserting $L_{t}= R_{t}^{1} N^{1}+ N^{3}K_{t}^{22} N^{2}$ into \eqref{p16}, we have
\begin{align*}
P_{t|t}&=(I-( R_{t}^{1} N^{1}+ N^{3}K_{t}^{22} N^{2})C)P_{t|t-1}\\
&\quad\times(I-( R_{t}^{1} N^{1}+ N^{3}K_{t}^{22} N^{2})C)\tr\\
&\quad+( R_{t}^{1} N^{1}+ N^{3}K_{t}^{22} N^{2})V( R_{t}^{1} N^{1}+ N^{3}K_{t}^{22} N^{2})\tr.
\end{align*}
Taking $ R_{t}^{1}$ as a variable, $\trace(P_{t|t})$ has the form
$\trace(P_{t|t})=\trace\Big( R_{t}^{1}( U_{t}^{3})^{-1}{ R_{t}^{1}}\tr+r_{t,R}\Big)$, where $r_{t,R}$ is a linear function  of $ R_{t}^{1}$.
Using the formula
$\trace(AXBX\tr)=\textmd{vec}\tr(X)(B\tr\otimes A)\textmd{vec}(X)$, one has
$\trace(P_{t|t})=\textmd{vec}\tr( R_{t}^{1})(( U_{t}^{3})^{-1}\otimes I)\textmd{vec}( R_{t}^{1})+\bar{r}_{t,R}$, where $\bar{r}_{t,R}$ is a linear function of $\textmd{vec}( R_{t}^{1})$.
Thus, $\frac{\partial^{2} \trace(P_{t|t})}{\partial^{2} \textmd{vec}( R_{t}^{1})}=( U_{t}^{3})^{-1}\otimes I\succ0$.
Similarly, taking $K_{t}^{22}$ as a variable, we have
 $\frac{\partial^{2} \trace(P_{t|t})}{\partial^{2} \textmd{vec}(K_{t}^{22})}=( U_{t}^{6})^{-1}\otimes { N^{3}}\tr N^{3}=( U_{t}^{6})^{-1}\otimes I\succ0$.
Thus, $\trace(P_{t|t})$ is convex with respect to both $ R_{t}^{1}$ and $K_{t}^{22}$. The optimal $ R_{t}^{1}$ and $K_{t}^{22}$ are given by
solving $\frac{\partial\trace(P_{t|t})}{\partial R_{t}^{1}}=0$ and $\frac{\partial\trace(P_{t|t})}{\partial K_{t}^{22}}=0$. That is
\begin{align*}
\frac{\partial\trace(P_{t|t})}{\partial R_{t}^{1}}&= R_{t}^{1}( U_{t}^{3})^{-1}
+ N^{3}K_{t}^{22} U_{t}^{1}+ U_{t}^{2}=0\\
\frac{\partial\trace(P_{t|t})}{\partial K_{t}^{22}}&=K_{t}^{22}( U_{t}^{6})^{-1}+{ N^{3}}\tr R_{t}^{1}
 U_{t}^{4}+{ N^{3}}\tr U_{t}^{5}=0,
\end{align*}
which gives
\begin{align*}
 R_{t}^{1*}=-( N^{3} J_{t}^{1} J_{t}^{2} U_{t}^{1}+ U_{t}^{2}) U_{t}^{3},\quad
K_{t}^{22*}
= J_{t}^{1} J_{t}^{2}.
\end{align*}
As a result, if $\gamma_{t,1}=0$, $\gamma_{t,2}=1$, then the optimal $L_{t}= R_{t}^{1*} N^{1}+ N^{3}K_{t}^{22*} N^{2}=L_{t}^{[01]}$.

(\textbf{Case 3}) If $\gamma_{t,1}=1$, $\gamma_{t,2}=0$, then $L_{t}$ is of the form $L_{t}=N^{4}K_{t}^{11}N^{1}+ R_{t}^{2}N^{2}$. Following from the the derivation similar to the one of \textbf{Case 2}, we have the optimal $K_{t}^{11}$ and $ R_{t}^{2}$ are given by
\begin{align*}
K_{t}^{11*}= J_{t}^{3} J_{t}^{4},\quad
 R_{t}^{2*}=-( N^{4} J_{t}^{3} J_{t}^{4} U_{t}^{4}+ U_{t}^{5}) U_{t}^{6}.
\end{align*}
Thus, if $\gamma_{t,1}=1$, $\gamma_{t,2}=0$, then the optimal $L_{t}= N^{4}K_{t}^{11*} N^{1}+ R_{t}^{2*} N^{2}=L_{t}^{[10]}$.

(\textbf{Case 4}) If $\gamma_{t,1}=0$, $\gamma_{t,2}=0$, then $L_{t}= N^{4}K_{t}^{11} N^{1}+ N^{3}K_{t}^{22} N^{2}$. Similarly, we obtain that the optimal $K_{t}^{11}$
and $K_{t}^{22}$ are of the form
\begin{align*}
K_{t}^{11*}&=-{ N^{4}}\tr( N^{3}K_{t}^{22} U_{t}^{1}+ U_{t}^{2}) U_{t}^{3}=-{ N^{4}}\tr U_{t}^{2} U_{t}^{3},\\
K_{t}^{22*}&=-{ N^{3}}\tr( N^{4}K_{t}^{11} U_{t}^{4}+ U_{t}^{5}) U_{t}^{6}=-{ N^{3}}\tr U_{t}^{5} U_{t}^{6}.
\end{align*}
Hence, if  $\gamma_{t,1}=0$, $\gamma_{t,2}=0$, the optimal $L_{t}= N^{4}K_{t}^{11*} N^{1}+ N^{3}K_{t}^{22*} N^{2}=L_{t}^{[00]}$. The proof is completed.
\subsection{The proof of Proposition \ref{lem1}}
\label{c7-1}
The equations \eqref{g1} are obvious. We focus on proving \eqref{g3}.

For any $Y^{1},Y^{2}\succeq 0$, let $Y=\alpha Y^{1}+(1-\alpha) Y^{2}$. One has
\begin{align*}
&\quad g_{\lambda_{1}\lambda_{2}}(Y)\\
&=  f( L^{[00]}[Y],L_{t}^{[01]}[Y], L^{[10]}[Y], L^{[11]}[Y],Y)\\
&=  f( L^{[00]}[Y], L^{[01]}[Y], L^{[10]}[Y], L^{[11]}[Y],\alpha Y^{1}+(1-\alpha) Y^{2})\\
&=\alpha  f( L^{[00]}[Y], L^{[01]}[Y], L^{[10]}[Y], L^{[11]}[Y],Y^{1})\\
&\quad+(1-\alpha)  f( L^{[00]}[Y], L^{[01]}[Y], L^{[10]}[Y], L^{[11]}[Y],Y^{2})\\
&\succeq\min_{S^{1}\in\Lambda^{[00]},S^{2}\in\Lambda^{[01]},S^{3}\in\Lambda^{[10]},S^{4}\in\Lambda^{[11]}}\alpha  f(S^{1},S^{2},S^{3},S^{4},Y^{1})\\
&\quad+\min_{S^{1}\in\Lambda^{[00]},S^{2}\in\Lambda^{[01]},S^{3}\in\Lambda^{[10]},S^{4}\in\Lambda^{[11]}}(1-\alpha)  \nonumber\\
&\quad\times f(S^{1},S^{2},S^{3},S^{4},Y^{2})\\
&=\alpha  f( L^{[00]}[Y^{1}], L^{[01]}[Y^{1}], L^{[10]}[Y^{1}], L^{[11]}[Y^{1}],Y^{1})\\
&\quad+(1-\alpha)  f( L^{[00]}[Y^{2}], L^{[01]}[Y^{2}], L^{[10]}[Y^{2}], L^{[11]}[Y^{2}],Y^{2})\\
&=\alpha g_{\lambda_{1}\lambda_{2}}(Y^{1})+(1-\alpha)g_{\lambda_{1}\lambda_{2}}(Y^{2}),
\end{align*}
where
$\Lambda^{[00]}=\Big\{\begin{bmatrix}
            X^{11} & 0 \\
            0 & X^{22} \\
          \end{bmatrix}:~X^{11}\in R^{n_{1}\times m_{1}},~
X^{22}\in R^{n_{2}\times m_{2}}
\Big\}$,
$\Lambda^{[01]}=\Big\{\begin{bmatrix}
            X^{11} & 0 \\
            X^{21} & X^{22} \\
          \end{bmatrix}:~X^{11}\in R^{n_{1}\times m_{1}}X^{21}\in R^{n_{2}\times m_{1}}, X^{22}\in R^{n_{2}\times m_{2}}
\Big\}$,
$\Lambda^{[10]}=\Big\{\begin{bmatrix}
            X^{11} & X^{12} \\
           0 & X^{22} \\
          \end{bmatrix}:~X^{11}\in R^{n_{1}\times m_{1}}X^{12}\in R^{n_{1}\times m_{2}}, X^{22}\in R^{n_{2}\times m_{2}}
\Big\}$, $\Lambda^{[11]}=\mathbb{R}^{n\times m}$.

This shows that $g_{\lambda_{1}\lambda_{2}}(Y)$ is concave. Using Jensen's Inequality, one has
$\ve(g_{\lambda_{1}\lambda_{2}}(Y))\preceq g _{\lambda_{1}\lambda_{2}}(\ve(Y))$. This completes the proof.

\subsection{The proof of Lemma \ref{elem2}}
(\emph{Sufficiency:})
Construct a sequence $\tilde{Y}_{0},~\tilde{Y}_{1},~\tilde{Y}_{2},\cdots$ by
\begin{align}
\tilde{Y}_{t+1}&= f\big( N^{4}X^{1} N^{1}+ N^{3}X^{2} N^{2},X^{3} N^{1}+ N^{3}X^{4} N^{2},\nonumber\\
&\quad \quad \quad  N^{4}X^{4} N^{1}+X^{6} N^{2},X^{7},\tilde{Y}_{t}\big), ~ \tilde{Y}_{0}=P_{0}.
\end{align}
Using the vectorization and Kronecker products, we can obtain that
$
\textmd{vec}(\tilde{Y}_{t+1})=  h(X^{1},\cdots,X^{7}) \textmd{vec}(\tilde{Y}_{t})+Q
$, where $Q$ is a bounded vector that does not depends on $\tilde{Y}_{t}$.
If $\rho\Big(  h(X^{1},\cdots,X^{7})\Big)<1$, $\lim\limits_{t\rightarrow +\infty}\textmd{vec}(\tilde{Y}_{t})$  is bounded, i.e. $\lim\limits_{t\rightarrow +\infty} \tilde{Y}_{t}$ is bounded. It follows from
\begin{align}
\label{e44}
&g_{\lambda_{1}\lambda_{2}}(Y)=\min\limits_{X^{1},\ldots,X^{7}}
 f\Big( N^{4}X^{1} N^{1}+ N^{3}X^{2} N^{2},X^{3} N^{1}\nonumber\\
&\quad+ N^{3}X^{4} N^{2}, N^{4}X^{4} N^{1}+X^{6} N^{2},X^{7},Y\Big),
\end{align}
that
$Y_{t}\preceq \tilde{Y}_{t}$. Hence, $\lim\limits_{t\rightarrow +\infty} Y_{t}$ is bounded.

(\emph{Necessity:}) From \eqref{e44}, the necessity is obvious. The proof is completed.

\subsection{The proof of Theorem \ref{zz2}}
\label{aap22c}
Using Schur complement for \eqref{e33} and \eqref{e38}, we have there exists $X^{1}$, $X^{2}$ such that $\Big(A-A( N^{4}X^{1} N^{1}+ N^{3}X^{2} N^{2}))C\Big)\Big(A-A( N^{4}X^{1} N^{1}+ N^{3}X^{2} N^{2}))C\Big)\tr \preceq r_{1} I$, which means that
$||A-A( N^{4}X^{1} N^{1}+ N^{3}X^{2} N^{2}))C||\leq\sqrt{r_{1}}$. It follows from the  formula $||A\otimes B||=||A||\times ||B||$ that
$|| d( N^{4}X^{1} N^{1}+ N^{3}X^{2} N^{2})||\leq r_{1}$. Similarly, there exists $X^{3},\ldots,X^{7}$ satisfying
$|| d(X^{3} N^{1}+ N^{3}X^{4} N^{2})||\leq r_{2}$,
$|| d( N^{4}X^{4} N^{1}+X^{6} N^{2})||\leq r_{3}$,
 $|| d(X^{7})||\leq r_{4}$.
As a result, there exists $X^{1},\ldots,X^{7}$ such that $\rho\Big(h(X^{1},\cdots,X^{7})\Big)\leq \lambda_{1}\lambda_{2}|| d( N^{4}X^{1} N^{1}+ N^{3}X^{2} N^{2})||+
\lambda_{1}(1-\lambda_{2})|| d(X^{3} N^{1}+ N^{3}X^{4} N^{2})||+(1-\lambda_{1})\lambda_{2}|| d( N^{4}X^{4} N^{1}+X^{6} N^{2})||
+(1-\lambda_{1})(1-\lambda_{2})|| d(X^{7})||\leq r_{1}\lambda_{1}\lambda_{2}+r_{2}\lambda_{1}(1-\lambda_{2})+r_{3}(1-\lambda_{1})\lambda_{2}+r_{4}(1-\lambda_{1})(1-\lambda_{2})< 1$. It follows from Lemma 2 that $\lim\limits_{t\rightarrow +\infty}Y_{t}$ is bounded. From Lemma 1, we know  $\lim\limits_{t\rightarrow +\infty}\ve(P_{t})\preceq \lim\limits_{t\rightarrow +\infty} Y_{t}$. Thus,
 $\lim\limits_{t\rightarrow +\infty}\ve(P_{t})$ is  bounded. The proof is completed.

\subsection{The proof of Corollary \ref{coro}}
Consider \eqref{zz33}. The optimal $X,\tilde{X}$ to the following optimization problem
\begin{align}
\min_{r,X,\bar{X}}\Big\{r\geq0:~ p_{3}(r,X,\tilde{X})\succeq 0\Big\}
\end{align}
is denoted by $X^{o3},\tilde{X}^{o3}$.
Then, we have
$ p_{3}(r_{3},X^{o3},\tilde{X}^{o3})= p(r_{3}, N^{4}X^{o3} N^{1}+\tilde{X}^{o3} N^{2})\succeq 0$. This implies that $r=r_{3}$, $X= N^{4}X^{o3} N^{1}+\tilde{X}^{o3} N^{2}$ is a solution to $ p(r,X)\succeq 0$.
It follows from the definition of $r_{4}$ (see \eqref{e36}) that $r_{4}\leq r_{3}$. Similarly, we can prove $r_{4}\leq r_{2}$, $r_{2}\leq r_{1}$, $r_{3}\leq r_{1}$.
If $r_{1}=r_{4}$, it follows from $r_{4}\leq r_{3}$, $r_{4}\leq r_{2}$, $r_{2}\leq r_{1}$, $r_{3}\leq r_{1}$ that $r_{1}=r_{2}=r_{3}=r_{4}$. According to Theorem 2, if $r_{1}=r_{2}=r_{3}=r_{4}< 1$, $\lim\limits_{t\rightarrow +\infty}\ve(P_{t})$  is bounded for any $\lambda_{1}$, $\lambda_{2}$.
The proof is completed.

\subsection{The proof of  Corollary \ref{coro2}}
From Corollary 1, we know that 1) $r_{4}\leq r_{1}$, and 2) if $r_{1}=r_{4}< 1$, then $\lim\limits_{t\rightarrow +\infty}\ve(P_{t})$ is  bounded for any $\lambda_{1}$, $\lambda_{2}\in[0~ 1]$.
 Now, we start to prove $r_{1}\neq r_{4}$ by contradiction.
According to the definition of $r_{4}$, one has  $r_{4}< 1$
 if there exists $X$ such that $c(X)\prec I$. Thus, $\lim\limits_{t\rightarrow +\infty}\ve(P_{t})$  is bounded for any $\lambda_{1}$, $\lambda_{2}$ if $c(X)\prec I$ and $r_{1}=r_{4}$ holds. However, under Assumption 1, $\lim\limits_{t\rightarrow +\infty}\ve(P_{t})$ is unbounded for $\lambda_{1}=\lambda_{2}=1$, because $(A,C)$ is undetectable with $\textmd{diag}\{1_{n_{1}\times m_{1}}, 1_{n_{2}\times m_{2}}\}$. This is a contradiction.  As a result,  $r_{1}\neq r_{4}$ if there exists $X$ such that
$c(X) \prec I$. The proof is completed.

\subsection{The proof of Lemma \ref{lee6}}
\label{app2}
To prove Lemma \ref{lee6}, we need Proposition 1. It is known that
Lemma \ref{lee6} holds if $\ve(P_{t})$ is monotone increasing with respect to $\lambda_{1},\lambda_{2}$.
From Proposition \ref{lem1}, one has  $\ve(P_{t+1})=\ve(g_{\lambda_{1}\lambda_{2}}(P_{t}))$. Hence, we only need to prove that $g_{\lambda_{1}\lambda_{2}}(Y)$ is monotone increasing with respect to $\lambda_{1},\lambda_{2}$ for any $Y\succ 0$. In particular, we should show that
\begin{itemize}
  \item If $0\leq\lambda_{1}\leq 1$ is fixed and $0\leq\lambda_{2}^{[1]}\leq\lambda_{2}^{[2]}\leq1$,
then $g_{\lambda_{1}\lambda_{2}^{[1]}}(Y)\preceq g_{\lambda_{1}\lambda_{2}^{[2]}}(Y)$.
  \item  If $0\leq\lambda_{2}\leq 1$ is fixed and $0\leq\lambda_{1}^{[1]}\leq\lambda_{1}^{[2]}\leq 1$, then $g_{\lambda^{[1]}_{1}\lambda_{2}}(Y)\preceq g_{\lambda_{1}^{[2]}\lambda_{2}}(Y)$.
\end{itemize}
From the definition of $g_{\lambda_{1}\lambda_{2}}(Y)$, one has
\begin{align*}
&\quad g_{\lambda_{1}\lambda_{2}^{[1]}}(Y)- g_{\lambda_{1}\lambda_{2}^{[2]}}(Y)\\
&=\lambda_{1}(\lambda_{2}^{[1]}-\lambda_{2}^{[2]})b(L_{t}^{00}[Y],Y)
+\lambda_{1}(\lambda_{2}^{[2]}-\lambda_{2}^{[1]})b( L^{[01]}[Y],Y)\\
&\quad+(1-\lambda_{1})(\lambda_{2}^{[1]}-\lambda_{2}^{[2]})b( L^{[10]}[Y],Y)\\
&\quad+(1-\lambda_{1})(\lambda_{2}^{[2]}-\lambda_{2}^{[1]})b( L^{[11]}[Y],Y)\\
&=(1-\lambda_{1})(\lambda_{2}^{[2]}-\lambda_{2}^{[1]})\Big[b( L^{[11]}[Y],Y)-b( L^{[10]}[Y],Y)\Big]\\
&\quad+\lambda_{1}(\lambda_{2}^{[2]}-\lambda_{2}^{[1]})\Big[b( L^{[01]}[Y],Y)-b( L^{[00]}[Y],Y)\Big].
\end{align*}
According to the definitions of $b(X,Y)$, and $ L^{\gamma}[Y]$, $\gamma\in\{[00],[01],[10],[11]\}$ (see after equation \eqref{zz15}), one has
$b( L^{[00]}[Y],Y)=\min\limits_{X\in\Lambda^{[00]}}b(X,Y)$,
$b( L^{[01]}[Y],Y)=\min\limits_{X\in\Lambda^{[01]}}b(X,Y)$,
$b( L^{[10]}[Y],Y)=\min\limits_{X\in\Lambda^{[10]}}b(X,Y)$,
$b( L^{[11]}[Y],Y)=\min\limits_{X\in\Lambda^{[11]}}b(X,Y)$,
where $\Lambda^{[00]},~\Lambda^{[01]},~\Lambda^{[10]},~\Lambda^{[11]}$ are defined in the proof of Proposition \ref{lem1}.
It follows from $\Lambda^{[11]}\supseteq\Lambda^{[10]}$, $\Lambda^{[01]}\supseteq\Lambda^{[00]}$ that
 $b( L^{[11]}[Y],Y)\preceq b( L^{[10]}[Y],Y)$, $b( L^{[01]}[Y],Y)\preceq b(L_{t}^{[00]}[Y],Y)$. As a result, $g_{\lambda_{1}\lambda_{2}^{[1]}}- g_{\lambda_{1}\lambda_{2}^{[2]}}\preceq0$. Similarly, we can obtain that $g_{\lambda_{1}^{[1]}\lambda_{2}}- g_{\lambda_{1}^{[2]}\lambda_{2}}\preceq0$. This completes the proof.

\subsection{The proof of Theorem \ref{co1}}
\label{app3}
Based on Lemma \ref{lee6}, we need to show two points: (\textbf{\textit{Point 1:}}) $\lim\limits_{t\rightarrow\infty}\ve(P_{t})$ is bounded if $\lambda_{2}\leq\underline{\lambda}_{2,c}$;
(\textbf{\textit{Point 2:}}) $\lim\limits_{t\rightarrow\infty}\ve(P_{t})$ is unbounded if $\lambda_{2}>\bar{\lambda}_{2,c}$.

\textbf{\textit{Point 1:}}
For the case $r_{1}=r_{4}\geq1$, $\underline{\lambda}_{2,c}=0$ is obtained directly from $0\leq\lambda_{2}\leq 1$. For the case $r_{1}=r_{4}<1$, according to Corollary 1, $\lim\limits_{t\rightarrow\infty}\ve(P_{t})$ is bounded  when $\lambda_{2}=1$. Thus, $\underline{\lambda}_{2,c}=\bar{\lambda}_{2,c}=1$ if $r_{1}=r_{4}<1$.
For the case $r_{1}\neq r_{4}$,
according to \eqref{ee5}, $\lambda_{2}\leq\underline{\lambda}_{2,c}$ becomes
\begin{align}
\label{z43}
\lambda_{2}\leq\frac{1-r_{2}\lambda_{1}-r_{4}(1-\lambda_{1})}{(r_{1}-r_{2})\lambda_{1}+(r_{3}-r_{4})(1-\lambda_{1})},
\end{align}
Since  $r_{4}\leq \min(r_{2},r_{3})$, $r_{1}\geq\max(r_{2},r_{3})$, and $r_{1}\neq r_{4}$, one has $r_{1}\neq r_{2}$ or $r_{3}\neq r_{4}$ holds. Thus,
$(r_{1}-r_{2})\lambda_{1}+(r_{3}-r_{4})(1-\lambda_{1})>0$. Applying \eqref{z43} to Theorem 2 gives that $\lim\limits_{t\rightarrow\infty}\ve(P_{t})$ is bounded.

\textbf{\textit{Point 2:}}
Define $s_{\lambda_{1}\lambda_{2}}(Y)=\lambda_{1}\lambda_{2}b( L^{[00]}[Y],Y)$.
The proof is divided into two steps. (\textbf{Step 1:}) to prove that $\ve[P_{t}]\succeq s^{t}_{\lambda_{1}\lambda_{2}}(P_{0})$, where 
the definition of the composite function $f^{t}(\cdot)$ is defined in ``Notations'' paragraph.
(\textbf{Step 2:}) to prove that  $\lim\limits_{t\rightarrow\infty}s^{t}_{\lambda_{1}\lambda_{2}}(P_{0})$ is unbounded
for $\lambda_{2}>\bar{\lambda}_{2,c}$.

\textbf{(Step 1:)}
Denote
$ z_{\gamma}(Y)=b( L^{\gamma}[Y],Y)$, $\gamma\in\{[00],[01],[10],[11]\}$,
and
define
\begin{align}
z_{\gamma_{t-1},\gamma_{t-2},\ldots,\gamma_{0}}(\cdot)&= z_{\gamma_{t-1}}( z_{\gamma_{t-2}}(\ldots  z_{\gamma_{0}}(\cdot))).
\end{align}
It follows from the  mathematical expectation formula, \eqref{zv11} and \eqref{c20} that
\begin{align}
\label{c52}
\ve[P_{t}]&=\sum_{\gamma_{0},\ldots,\gamma_{t-1}\in\{11,01,10,00\}}\pr(\gamma_{0},\cdots,\gamma_{t-1})\nonumber\\
&\quad\quad\quad\quad\quad\quad\times z_{\gamma_{t-1},\gamma_{t-2},\ldots,\gamma_{0}}(P_{0}).
\end{align}
Because $\gamma_{t}=[\gamma_{t,1}~\gamma_{t,2}]$ is an i.i.d. process and satisfies \eqref{zv3}, one has
\begin{align}
\label{c53}
\pr(\gamma_{0},\cdots,\gamma_{t-1})=\pr(\gamma_{0})\pr(\gamma_{1})\cdots\pr(\gamma_{t-1}),
\end{align}
where for any $i\in \{0,\ldots,t-1\}$,
\begin{align*}
\pr(\gamma_{i}=[11])&=(1-\lambda_{1})(1-\lambda_{2}),\pr(\gamma_{i}=[01])=(1-\lambda_{1})\lambda_{2},\\
\pr(\gamma_{i}=[10])&=\lambda_{1}(1-\lambda_{2}),\pr(\gamma_{i}=[00])=\lambda_{1}\lambda_{2}.
\end{align*}
It is known that $s^{t}_{\lambda_{1}\lambda_{2}}(P_{0})=(\lambda_{1}\lambda_{2})^{t} z_{[00]}^{t}(P_{0})$.
From \eqref{c52} and \eqref{c53}, we can obtain
$\ve[P_{t}]=h^{t}_{\lambda_{1}\lambda_{2}}(P_{0})+\Gamma$, where $\Gamma\succeq0$, because $ z_{\gamma_{t-1}\gamma_{t-2}\ldots \gamma_{0}}(P_{0})\succeq0$ holds for any
$\gamma_{0},\ldots,\gamma_{t-1}\in\{[11],[01],[10],[00]\}$. Thus, we have $s^{t}_{\lambda_{1}\lambda_{2}}(P_{0})\preceq\ve[P_{t}]$.

\textbf{(Step 2:)}
If $\lambda_{2}>\bar{\lambda}_{2,c}$, then $\lambda_{1}\lambda_{2}\underline{\delta}\Big(\min\limits_{X^{1},X^{2}}q(X^{1},X^{2})\Big)>1$.
Denote
\begin{align*}
&l(Y)=\lambda_{1}\lambda_{2}a( L^{[00]}[Y],Y),\\
&\hat{l}(X^{1},X^{2},Y)=\lambda_{1}\lambda_{2}a( N^{4}X^{1} N^{1}+ N^{3}X^{2} N^{2},Y).
\end{align*}
%
From
$s_{\lambda_{1}\lambda_{2}}(Y)\succ l(Y)=\min\limits_{X^{1},X^{2}}\hat{l}(X^{1},X^{2},Y)$, and the fact that both $s_{\lambda_{1}\lambda_{2}}(Y)$ and $\min\limits_{X_{1},X_{2}}\hat{l}(X^{1},X^{2},Y)$ are monotonically increasing with respect to $Y$, one has
$s^{t}_{\lambda_{1}\lambda_{2}}(P_{0})\succ \breve{l}^{t}(P_{0})$, where $\breve{l}(P_{0})=
\min\limits_{X^{1},X^{2}}\hat{l}(X^{1},X^{2},P_{0})$.

It is known that
$\trace\big(\breve{l}(P_{0})\big)
=\trace\big(\min\limits_{X^{1},X^{2}}\lambda_{1}\lambda_{2}q(X^{1},X^{2})P_{0}\big)
$,
where $q(X^{1},X^{2})$ is defined before Theorem \ref{co1}.
Because $\lambda_{1}\lambda_{2}\underline{ \delta}\Big(\min\limits_{X^{1},X^{2}}q(X^{1},X^{2})\Big)>1$, one can obtain $\trace\big(\breve{l}^{t}(P_{0})\big)\rightarrow\infty$
when $t\rightarrow\infty$. Recall that $s^{t}_{\lambda_{1}\lambda_{2}}(P_{0})\succ\breve{l}^{t}(P_{0})$. Thus,
$\lim\limits_{t\rightarrow\infty}s^{t}_{\lambda_{1}\lambda_{2}}(P_{0})$ is unbounded for $\lambda_{2}>\bar{\lambda}_{2,c}$. The proof is completed.

\begin{IEEEbiography}[{
\includegraphics[width=1in,height=1.25in,clip,keepaspectratio]{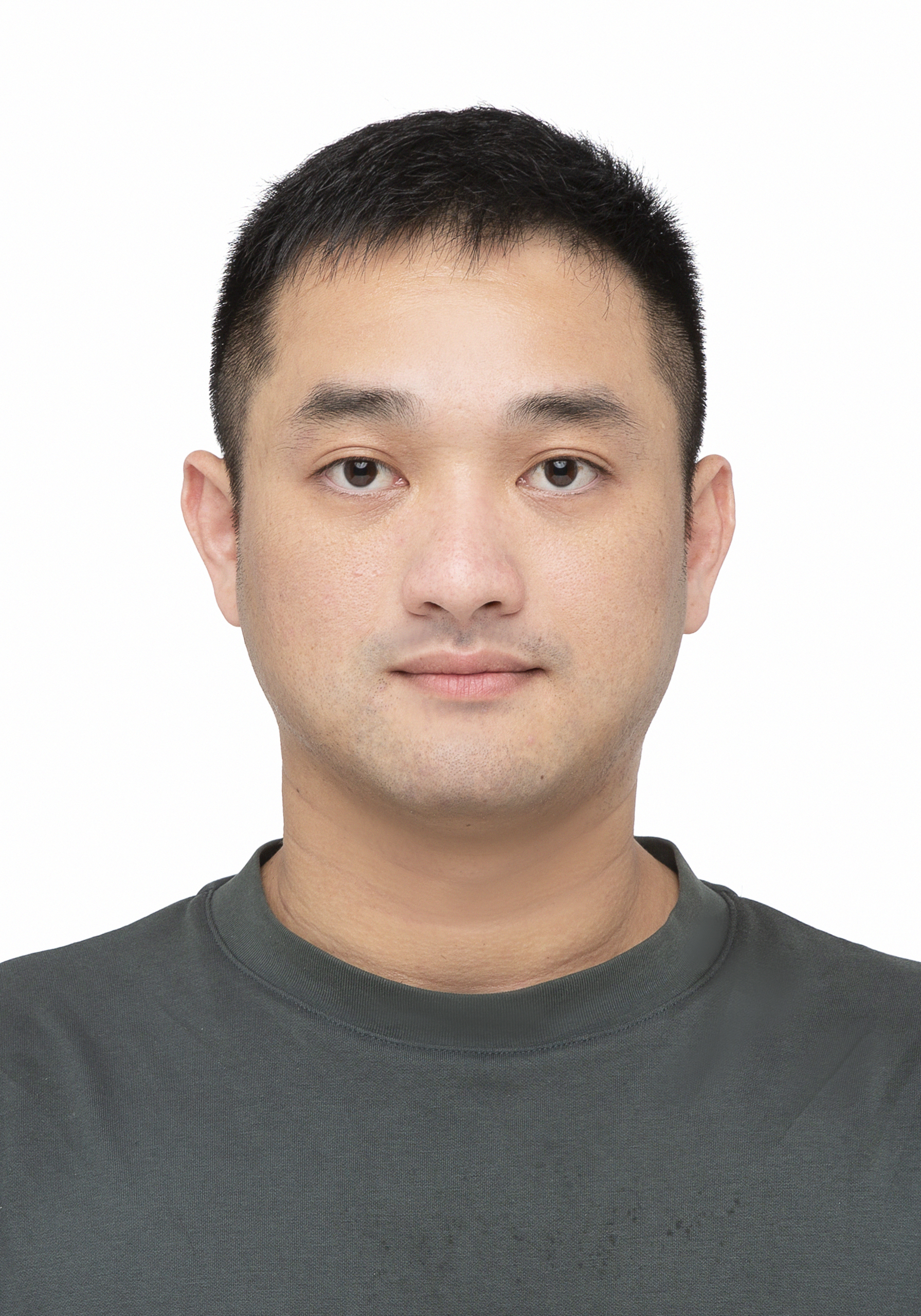}}]{}
\noindent{\bf Yan Wang}~ obtained the B.E. degree in automation and the Ph.D. degree in control science and engineering from University of Science and Technology of China in 2014 and 2019, respectively. During 2019 to 2021, he was a Research Fellow at Nanyang Technological University, Singapore. From 2021 to Jan. 2023, he was affiliated with The Hong Kong Polytechnic University, and Chinese University of Hong Kong. He is currently an Associate Professor at the School of Mechanical and Electrical Engineering and Automation, Harbin Institute of Technology Shenzhen, Shenzhen, China. He¡¯s research interests include optimal control/estimation of interconnected systems; connected vehicle system control and optimization; AGV schedule in flexible manufacturing systems.
\end{IEEEbiography}

\begin{IEEEbiography}[{
\includegraphics[width=1in,height=1.25in,clip,keepaspectratio]{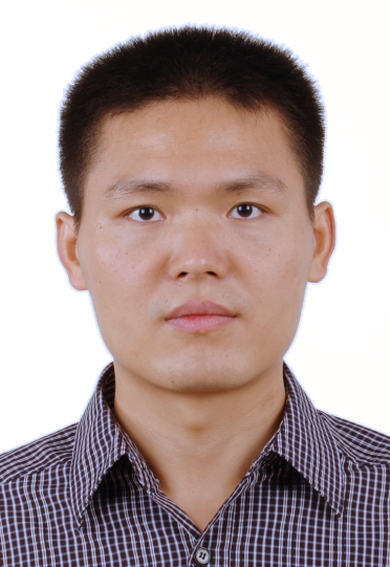}}]{}
\noindent{\bf Junlin Xiong}~ received his BEng. and MSci. degrees from Northeastern University, China, and his PhD degree from the University of Hong Kong, Hong Kong, in 2000, 2003 and 2007, respectively. From 2007 to 2010, he was a research associate at the University of New South Wales at the Australian Defence Force Academy, Australia. In March 2010, he joined the University of Science and Technology of China where he is currently a professor in the Department of Automation. Currently, he is an Associate Editor for the IET Control Theory and Application. His current research interests are in the fields of negative imaginary systems, large-scale systems and networked control systems.
\end{IEEEbiography}

\begin{IEEEbiography}[{
\includegraphics[width=1in,height=1.25in,clip,keepaspectratio]{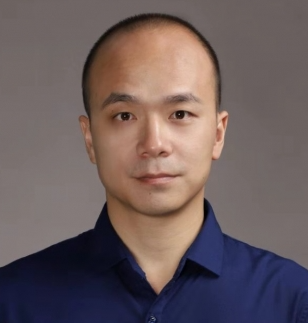}}]{}
\noindent{\bf Zaiyue Yang }~received the B.S. and M.S. degrees from the Department of Automation, University of Science and Technology of China, Hefei, China, in 2001 and 2004, respectively, and the Ph.D. degree from the Department of Mechanical Engineering, University of Hong Kong in 2008. He was a Postdoctoral Fellow and a Research Associate with the Department of Applied Mathematics, Hong Kong Polytechnic University before joining the College of Control Science and Engineering, Zhejiang University, Hangzhou, China, in 2010. Then, he joined the Department of Mechanical and Energy Engineering, Southern University of Science and Technology, Shenzhen, China, in 2017, where he is currently a Professor. His current research interests include smart grid, signal processing, and control theory. He is an Associate Editor for the IEEE Transactions on Industrial Informatics.
\end{IEEEbiography}

\begin{IEEEbiography}[{
\includegraphics[width=1in,height=1.25in,clip,keepaspectratio]{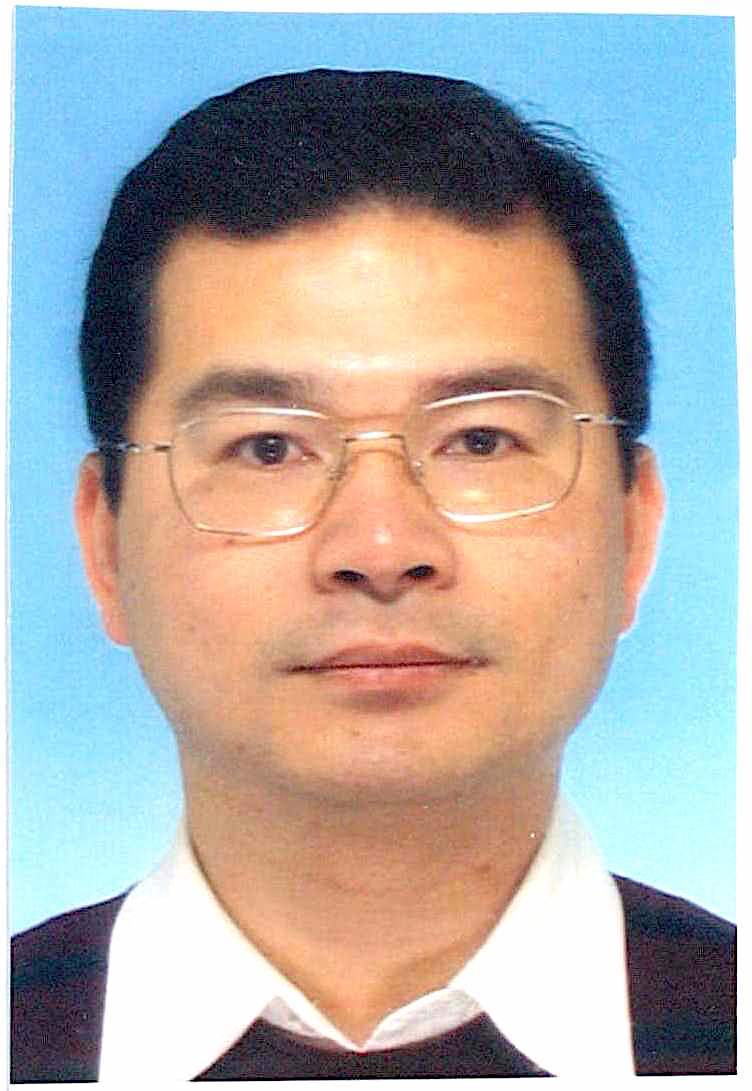}}]{}
\noindent{\bf Rong Su }~received the Bachelor of Engineering degree
from University of Science and Technology of China in
1997, and the Master of Applied Science degree and
Ph.D. degree from University of Toronto, in 2000 and
2004, respectively. He was affiliated with University of
Waterloo and Technical University of Eindhoven before
he joined Nanyang Technological University in 2010.
Currently, he is an associate professor in the School of
Electrical and Electronic Engineering. Dr. Su¡¯s research
interests include multi-agent systems, cybersecurity
of discrete-event systems, supervisory control, model-based fault diagnosis, control and optimization in complex networked systems
with applications in flexible manufacturing, intelligent transportation, human¨C
robot interface, power management and green buildings. In the aforementioned
areas he has more than 220 journal and conference publications, and 5 granted
USA/Singapore patents. Dr. Su is a senior member of IEEE, and an associate
editor for Automatica, Journal of Discrete Event Dynamic Systems: Theory
and Applications, and Journal of Control and Decision. He was the chair of
the Technical Committee on Smart Cities in the IEEE Control Systems Society
in 2016¨C2019, and is currently the chair of IEEE Control Systems Chapter,
Singapore

\end{IEEEbiography}

\end{document}